\documentclass[a4paper, 11pt]{article}
\usepackage{amsmath,amssymb,amsthm,amsfonts,epsf}
\usepackage{graphicx,url}
\usepackage{latexsym,framed,color}
\usepackage[margin=2cm]{geometry}
\usepackage{times}

\newcommand{\bfi}{\bfseries\itshape}

\newcommand{\rem}[1]{}

\newtheorem{theorem}{Theorem}

\newtheorem{remark}{Remark}
\makeatletter

\@addtoreset{figure}{section}
\def\thefigure{\thesection.\@arabic\c@figure}
\def\fps@figure{h, t}
\@addtoreset{table}{bsection}
\def\thetable{\thesection.\@arabic\c@table}
\def\fps@table{h, t}
\@addtoreset{equation}{section}

\makeatother

\begin{document}

\begin{center}

{\LARGE\textbf{Multi-component generalizations of the {CH} equation:\\
Geometrical Aspects, Peakons and Numerical Examples
\\}} \vspace {10mm} \vspace{1mm} \noindent

{{\bf D. D. Holm $^{1}$} and  {\bf R. I.
Ivanov}\footnote{Department of Mathematics, Imperial College
London. London SW7 2AZ, UK.  \texttt{d.holm@imperial.ac.uk, r.ivanov@imperial.ac.uk}}
$^{,}$}\footnote{School of Mathematical Sciences, Dublin Institute of Technology, Kevin Street, Dublin 8, Ireland,  \texttt{rivanov@dit.ie}}

\vskip1cm \hskip-.3cm

\end{center}



\begin{abstract}
The Lax pair formulation of the two-component Camassa-Holm equation (CH2) is generalized to produce an integrable multi-component family, CH(n,k), of equations with $n$ components and $1\le |k|\le n$ velocities.  All of the members of the CH(n,k) family show fluid-dynamics properties with coherent solitons following particle characteristics. We determine their Lie-Poisson Hamiltonian structures and give numerical examples of their soliton solution behaviour. We concentrate on the CH(2,k) family with one or two velocities, including the CH(2,-1) equation in the Dym position of the CH2 hierarchy. A brief discussion of the CH(3,1) system reveals the underlying graded Lie-algebraic structure of the Hamiltonian formulation for CH(n,k) when $n\ge3$.
  



\end{abstract}



\section{Introduction}

{\bf Purpose.} \\
This paper develops the  Lax formulation for an integrable family of equations that contains the multi-component generalizations CH(n,k)  of the {CH} equation with $n$ momentum components convected by $1\le |{k}|\le {n}$ velocities. 
In the CH(n,k) family, CH is designated by CH(1,1) and CH2 by CH(2,1).
We also consider the Lie-Poisson Hamiltonian properties of several other members of the CH(n,k) family, particularly CH(2,1), CH(2,2), CH(2,-1), CH(3,1) and CH(3,2). The CH(2,-1) system may be regarded as a two-component generalization for CH of the Dym equation in the CH hierarchy. All the CH(n,k) equations for a given value of ${n}$ share the same spectral problem and therefore belong to the same integrable hierarchy. As might be expected, higher-order powers of the spectral parameter appear in the Lax formulation of CH(n,k) for greater values of ${n}$.  When compared to rigid rotations, the CH(2,1) shallow water system recovers the heavy top equations and the CH(2,2) system recovers the equations for a rigid body in a potential field. The CH(3,1) system reveals the underlying graded Lie-algebraic structure of the Hamiltonian formulation for CH(n,k). Examples of numerical solutions of these integrable systems of equations illustrate their interesting dynamical properties, in which soliton trains emerge from spatially confined initial conditions and interact with each other in a variety of different ways for the various CH(n,k) systems investigated here. Many open problems arise and we attempt to sketch some of the opportunities for future research in the conclusion section.

\subsection{Brief review of the Camassa-Holm (CH) equation}

The CH equation \cite{CH93, CHH94}
\begin{equation}\label{eq1}
 u_{t}-u_{xxt}+2\omega u_{x}+3uu_{x}-2u_{x}u_{xx}-uu_{xxx}=0,
\end{equation}
governs the evolution of the function $u(x,t): \mathbb{R}\times\mathbb{R}\to\mathbb{R}$, interpreted as a shallow water fluid velocity. When the linear dispersion parameter $\omega\in \mathbb{R}$ vanishes ($\omega=0$) the CH equation (\ref{eq1}) admits {\bfi peakon} solutions. Peakons are {\it nonanalytic} solitons that superpose as 
\begin{equation}\label{peakontrain-soln}
u(x,t)=\tfrac12\sum_{a=1}^Np_a(t)e^{-|x-q_a(t)|}
\,,
\end{equation}
for sets $\{p(t)\}$ and $\{q(t)\}$ satisfying a system of completely integrable canonical Hamiltonian equations.

For $\omega\ne0$, the CH equation (\ref{eq1}) describes the unidirectional propagation of dispersive shallow water waves over a flat bottom at one order higher than KdV in the standard asymptotic expansion of the Euler fluid equations with a free surface in a certain Galilean frame (mean wave speed) \cite{CH93,CHH94, DGH03, DGH04, J02, J03a}.  It also describes axially symmetric waves in a hyperelastic rod \cite{Dai98}. The inverse scattering problem for CH is treated, e.g., in \cite{dMonv,C01,CGI,CI06}. For a brief history of the developments and results about the CH equation of relevance in the present paper, one may consult the recent review \cite{HI10} and references therein.

The scope of the variety of mathematical interpretations of CH may be gleaned by rewriting it in various equivalent forms, many of which have been discovered several times before and each of which can be a point of departure for further investigation. For example, CH may be treated variously as: 
a fluid motion equation; a mathematical model of shallow water wave breaking; 
a vanishing Lie derivative, describing invariance of a 1-form density under the flow of a vector field related to it by inversion of the Helmholtz operator; 
a nonlocal characteristic equation; 
an Euler-Poincar\'e equation describing geodesic motion on the diffeomorphism group with respect to the metric defined by the $H^1$ norm on the tangent space of vector fields; 
a Lie-Poisson Hamiltonian system describing coadjoint motion on the Bott-Virasoro Lie group; 
a bi-Hamiltonian system; 
a compatibility equation for a linear system of two equations in a Lax pair, etc. 

Our own point of departure and primary emphasis in this paper is in treating CH as a member of a family of integrable evolutionary equations associated with a certain class of energy-dependent isospectral eigenvalue problems of polynomial order. We shall also discuss the geometrical interpretations and varieties of solution behaviour of several members of the hierarchy. To establish notation, we begin by writing CH in several of its various forms and discussing its properties that are relevant here. Most of these properties were already established  in \cite{CH93, CHH94}. 

\paragraph{CH as a fluid motion equation.}
CH may be written in the form of a {\bfi fluid motion equation}, as
\begin{eqnarray}  
u_t + uu_x = - P_x
\,,
\end{eqnarray}  
with {\bfi pressure} $P$ given by the convolution
\begin{eqnarray}   
P = K*(u^2 + \tfrac12 u_x^2 + \omega u)
\quad\hbox{with kernel}\quad
K(x,y)=\tfrac12\exp(-|x-y|).
\label{kernelK}
\end{eqnarray}  
The kernel $K$ is the Green's function for the 1D Helmholtz operator, $(1-\partial_x^2)$, and is also the shape of the peakon profile in equation (\ref{peakontrain-soln}).

\paragraph{Lie derivative, or characteristic form of CH.}
Upon introducing a {\bfi momentum} variable
 \begin{eqnarray} 
 \label{eq4a} m = u-u_{xx} = (1-\partial_x^2)u, \end{eqnarray}  
CH expresses the {\bfi vanishing Lie derivative} condition for invariance of its momentum, as a 1-form density, along characteristics of an associated velocity vector field,
\begin{eqnarray} 
(\partial_t + \mathcal{L}_u)\big((m+\omega)\,dx^2 \big)
=
\Big(m_t + um_x + 2(m+\omega)u_x\Big)\,dx^2
=0
\,,\quad\hbox{with}\quad
u(x,t)=K*m
\,.
\label{Eul-inv}
\end{eqnarray} 
This form of CH may be interpreted as the the condition for the 1-form density $(m+\omega)\,dx^2$ to be preserved (frozen in) under the flow of a characteristic velocity $dx/dt=u(x(t),t)$, namely,
\begin{eqnarray} 
\frac{d}{dt}
\Big((m+\omega)\,dx^2 \Big) = 0
\,,\quad\hbox{along}\quad
\frac{dx}{dt}=u(x(t),t)=K*m
.
\label{Lag-inv}
\end{eqnarray} 
In the parlance of fluids, equation (\ref{Eul-inv}) is the {\bfi Eulerian} form of the invariance, while equation (\ref{Lag-inv}) is its equivalent {\bfi Lagrangian} form.  As with the evolution of vorticity according to Euler's equations for  incompressible fluid motion, the relation between the velocity of the flow and the property it carries is \emph{nonlocal}. For the Euler fluid equations, the velocity carries the vorticity, related to the velocity by the convolution in  the Biot-Savart law, which inverts the curl operation. For CH, the momentum is related to the velocity by convolution with the kernel $K$ in (\ref{kernelK}), which inverts the Helmholtz operator. When expressed in terms of the momentum, the peakon velocity solution
(\ref{peakontrain-soln}) of dispersionless CH for $\omega=0$ becomes a sum over delta
functions, supported on a set of points moving on the real line. That is, 
the peakon velocity solution (\ref{peakontrain-soln}) implies
\begin{equation}\label{CHpeakon-m-soln}
m(x,t)=\sum_{a=1}^N \, p_a(t)\delta(x-q_a(t))
\,,\end{equation}
because of the relation
$(1-\partial_x^2)(\frac12e^{-|x-y|})=\delta(x-y)$ for the kernel $K$ in (\ref{kernelK}).
As shown in \cite{HoMa2004}, the CH peakon solution (\ref{CHpeakon-m-soln}) is geometrically the  cotangent-lift momentum map for the left action of the diffeomorphisms ${\rm Diff}(\mathbb{R})$ on a set of $N$ points on the real line. 

\paragraph{The Euler-Poincar\'e and Lie-Poisson properties of CH}
The geometric properties of CH in  (\ref{eq1}) follow from Hamilton's principle with a Lagrangian $l(u):\, \mathfrak{X}(\mathbb{R})\to \mathbb{R}$ by using the Euler-Poincar\'e theory \cite{HoMaRa1998}. Its associated Lie-Poisson Hamiltonian formulation in terms of $m\in \mathfrak{X}^*(\mathbb{R})$ then emerges from a Legendre transformation. 
Here, the velocity vector field $u\in \mathfrak{X}(\mathbb{R})\simeq  T{\rm Diff}(\mathbb{R})/{\rm Diff}(\mathbb{R})$ in the Lagrangian $l(u)$ is right-invariant under ${\rm Diff}(\mathbb{R})$, the diffeomorphisms of the real line.  The Lagrangian for CH is $l(u)=\frac12\|u\|^2_{H^1}$, the $H^1$ norm on these vector fields; which provides the metric for the interpretation of CH solutions as geodesic motion on ${\rm Diff}(\mathbb{R})$. The variational derivative $\delta l/\delta u \in \mathfrak{X}^*(\mathbb{R})$, in the space of real-valued 1-form densities dual to $\mathfrak{X}(\mathbb{R})$ yields the CH momentum, 
\[
m = \frac{\delta l}{\delta u} = u - u_{xx}
\,.
\]
The Lie-Poisson Hamiltonian formulation of CH follows by Legendre transforming the Euler-Poincar\'e equation for right-invariant vector fields, as
\begin{equation}\label{EPLPstructure}
\frac{d}{dt}\frac{\delta l}{\delta u} =-\, {\rm ad}^*_u\frac{\delta l}{\delta u}
\quad\Longrightarrow\quad
m_t = -\, (\partial_x m +  m \partial_x )\frac{\delta h}{\delta m}
\end{equation}
with
\[
h(m) = \langle m,u\rangle - l(u)
\,,\quad
\frac{\delta h}{\delta m}=u
\,,\quad
\frac{\delta h}{\delta u}=0 = m - \frac{\delta l}{\delta u}
\,,
\]
where $\langle \,\cdot\,,\,\cdot\,\rangle:\mathfrak{X}^*(\mathbb{R})\times\mathfrak{X}(\mathbb{R})\to\mathbb{R}$ denotes $L^2$ pairing on the real line. For more detail of the Lie-Poisson structure of CH, see  \cite{HoMaRa1998}. For its interpretation as coadjoint motion on the Bott-Virasoro Lie group, see  \cite{M98}. The Euler-Poincar\'e and Lie-Poisson structure of the CH equation in (\ref{EPLPstructure}) makes it clear how to generalize it to higher dimensions. The result is the {\bfi EPDiff equation} in $n$ dimensions, given for $\omega=0$ by
\begin{equation}\label{EPDiff-eqn}
\partial_t m_i = -\, (\partial_j m_i +  m_j \partial_i)u^j
\quad\hbox{with velocity components} \quad
\frac{\delta h}{\delta m_j} = u^j
,
\end{equation}
where $i,j = 1,2,\dots, n$.

\paragraph{The bi-Hamiltonian property of CH.}
The one-dimensional  CH equation (\ref{eq1}) may be written in bi-Hamiltonian form as
\begin{equation}\label{eq2}
 m_{t}=-(\partial_x -\partial_x ^{3})\frac{\delta H_{2}[m]}{\delta m}
 =
 -\Big(\partial_x  (m+ \omega ) + (m + \omega)\partial_x \Big)\frac{\delta H_{1}[m]}{\delta m}\,,
\end{equation}

where the two Hamiltonians are given by
\begin{eqnarray}  \label{eq2a} H_{1}[m]
=
\tfrac{1}{2}\int m u\, dx
\qquad\hbox{and}\qquad
H_{2}[m]
=
\tfrac{1}{2}\int(u^{3}+uu_{x}^{2}+2\omega u^{2})dx\,. \end{eqnarray} 

The integration is over the real line, for functions that decay sufficiently rapidly as $|x|\to  \infty $, and over one period, for periodic functions. By Magri's theorem \cite{Magri1978}, the bi-Hamiltonian property of CH implies an infinite sequence of conservation laws, obtained by a recursion relation.

\paragraph{The Lax pair for CH.}
Its bi-Hamiltonian property also implies that the CH equation (\ref{eq1}) admits a {\bfi Lax pair representation}, given by \cite{CH93, CHH94}
\begin{eqnarray}  \label{eq3} \Psi_{xx}&=&\Big(\frac{1}{4}+\lambda
(m+\omega)\Big)\Psi
 ,\\\label{eq4}
\Psi_{t}&=&\Big(\frac{1}{2\lambda}-u\Big)\Psi_{x}+\frac{u_{x}}{2}\Psi+\gamma\Psi
,\end{eqnarray}

where $\omega,\gamma$ are arbitrary real constants and the eigenvalue $\lambda$ is independent of time. The compatibility of the Lax pair for constant $\lambda$ means that the eigenvalue equation in (\ref{eq3}) is {\bfi isospectral}. That is, its spectrum is invariant under the flow of the CH equation. 

Many papers have been written to explore various features of the original single-component CH equation. A brief history of its exploration is recounted, for example, in \cite{HI10}.  See also \cite{Ho2010} for a recent discussion of its singular peakon solutions.

\subsection*{Plan of the paper}

Section \ref{CHn-general} continues the introduction of our subject by discussing the extension of CH to the integrable CH(n,k) systems consisting of  $n$ components (momentum densities) and $1\le |
{k}|\le n$ velocities that result from the isospectral problem in (\ref{L1}) and (\ref{L2}). The CH(n,k)  hierarchy may be written in a compact universal form that aids in the physical interpretation of its various equations as continuum flows and is reminiscent of the Virasoro structure of the one-component CH equation found in \cite{M98}. 
Section \ref{examples CH2hierarchy} reviews the properties of the two-component CH(2,1) system in  (\ref{CH2-q})-(\ref{CH2-rho}) and then discusses the CH(2,2) system, both of which are fluid systems. 
Section \ref{eqns-hierarchy-sec} provides two other examples of equations in the CH2 hierarchy: (i) the CH(2,\,-\,1) system, in the position of the CH hierarchy corresponding to the Dym equation in the CH hierarchy; and (ii) the CH(2,1) system with two time variables. 
Section \ref{conclusion-sec} closes the paper by giving a brief summary of its main points and indicating some open problems for future research.

\section{A hierarchy of multi-component integrable extensions of CH}   \label{CHn-general}

The main feature of the Inverse Scattering Transform (IST) for the two-component generalization of the {CH} equation (CH2) is that its spectral problem is Schr\"odinger's equation with an `energy dependent' potential, in which higher-order powers of the spectral parameter appear. For the history and development of the IST method with energy dependent potentials, one may consult \cite{K75,JJ72,SS97,AFL91,MA02} and the references therein.

A hierarchy of multi-component generalizations of CH may be obtained by considering an extension of the Lax pair (\ref{eq3}), (\ref{eq4}) that preserves its form, but replaces its coefficients by polynomials in the scattering parameter $\lambda$, as in \cite{I06},  
\begin{eqnarray}  \label{L1} \Psi_{xx}&=&Q(x,\lambda)\Psi,
 \\\label{L2}
\Psi_{t}&=&-U(x,\lambda)\Psi_{x}+\frac{1}{2}U_x(x,\lambda)\Psi, \end{eqnarray} 
where the potential $Q(x,\lambda)$ has the following energy dependence
\begin{eqnarray}  \label{L3} Q(x,\lambda)&=&\lambda^n q_n(x)+\lambda^{n-1}
q_{n-1}(x)+\ldots+\lambda q_1(x)+\frac{1}{4},
 \\\label{L4}
U(x,\lambda)&=&u_0(x)+\frac{u_1(x)}{\lambda}+\ldots
\frac{u_k(x)}{\lambda^k}. \label{L4U} \end{eqnarray} 
The compatibility condition for (\ref{L1}), (\ref{L2}) gives the following
equation, 
\begin{eqnarray}  \label{L5} 
Q_t+(\partial_x Q + Q\partial_x)U = \frac{1}{2} U_{xxx},
\end{eqnarray} 
whose form is reminiscent of the Virasoro structure of the one-component CH equation found in \cite{M98}.

Following this important clue to the nature of the equations in this hierarchy, we rewrite the equation geometrically as the Lie derivative $\mathcal{L}_U$ of the 1-form density $(Q\,\text{d}x^2)$ with respect to the vector field $U$ as
\begin{eqnarray} 
(\partial_t + \mathcal{L}_U)(Q\,\text{d}x^2) = \frac{1}{2}( \text{d}U_{xx})dx,
\label{LieDerivForm}
\end{eqnarray} 
whose relation to continuum flows may be emphasized by rewriting it in characteristic form as
\begin{eqnarray} 
\frac{d}{dt}\Big(Q(x(t),t)\,\text{d}x(t)^2 \Big) = \frac{1}{2}( \text{d}U_{xx})dx(t),
\quad\hbox{along}\quad
\frac{dx}{dt} = U(x(t),t)
.
\label{CharacteristicForm}
\end{eqnarray} 
{\bf Interpretation.} The left hand side of equations (\ref{LieDerivForm}) and (\ref {CharacteristicForm}) represents sweeping of the wave momentum density $Q$ by the flow of the velocity vector field $U$, while the right hand side represents dispersion of the wave and forces that cause motion relative to the flow lines of $U$. 

{\bf CH(n,k) family of equations.} Upon substituting the expansions in $\lambda$ and $\lambda^{-1}$ from (\ref{L3}) and (\ref{L4}), respectively, into equation (\ref{L5}), one obtains a chain
of $n$ evolution equations with ${k}+1$ differential relations for
the $n+{k}+1$ variables $q_1$, $q_2$, $\ldots$, $q_n$, $u_0$, $u_1$,
$\ldots$, $u_k$ ($n$ and ${k}$ are arbitrary positive, or negative, integers):
%
\begin{eqnarray}   
q_{n-r,t}
&=&
-\sum_{s=\max(0,r-k)} ^{r}\big(\partial_xq_{n-s}+ q_{n-s}\partial_x\big)u_{r-s}\,,
\qquad r=0,1,\ldots,n-1
, \nonumber \\
0
&=&
\frac{1}{2}\big(\partial_x-\partial_x^3\big)u_r\
+
\sum_{s=1} ^{\min(n,k-r)}\big(\partial_xq_{s}+ q_{s}\partial_x\big)u_{r+s}
\,, \label{L6} \\
&\phantom{=}& \phantom{********************}  r=0,1,\ldots,k-1, \nonumber\\
 0&=&\big(\partial_x-\partial_x^3\big)u_k. \nonumber  \end{eqnarray} 
The differential relations in the middle equation of (\ref{L6}) contain the same (Hamiltonian) operators as in the biHamiltonian structure (\ref{eq2}). 

System (\ref{L6}) is similar to the hydrodynamic chain
studied  in a series of papers \cite{SMA,MA02,MA03}, and to
other CH generalizations \cite{GH03,HQ03,EP05,KZ00,CGP97}. 
In the present paper, we shall examine the geometric structure and numerical solution behavior of several examples of equations from the CH(n,k) family of equations  resulting from the isospectral problem (\ref{L1}) and (\ref{L2}).

\subsection{CH(n,1) system}

The simplest chain system with $n$ coupled equations arises from the general framework (\ref{L6}) for $k=1$. This system is denoted CH(n,1). In this case, the Lax pair (\ref{L1}) - (\ref{L2}) becomes 
\begin{eqnarray}
Q(x,\lambda)&=&-\lambda^n \rho^2 + \lambda^{n-1}
q_{n-1}+\ldots+\lambda q_{1} +\frac{1}{4}
\label{CH21-LPairQ}\,,\\
U(x,\lambda)&=&-\frac{1}{2\lambda}+u
\,.
\label{CH21-LPairU}
\end{eqnarray}
There is one differential relation, $q_1=u-u_{xx}+\text{const.}$ Compatibility of the corresponding Lax pair in this case results in a chain of equations in the following form for $p=1,2,\dots,n$,
\begin{eqnarray} \label{CH-n1}
\partial_t q_p + ( \partial_x q_p  + q_p \partial_x)u
-\tfrac{1}{2}\partial_x q_{p+1}=0 \quad \hbox{with} \quad q_n=-\rho^2 \quad \hbox{and} \quad q_{n+1}=0  .
\end{eqnarray} 

\subsection{CH(n,2) system}

The Lax pair (\ref{L1}) - (\ref{L2}) with
\begin{eqnarray}
Q&=&\lambda^n q_n + \lambda^{n-1}
q_{n-1}+\ldots+\lambda q_{1} +\frac{1}{4}
\,,\label{Qexp}\\
U&=&-\frac{1}{2\lambda^2}+\frac{u_1}{\lambda}+u_0
\,,\label{Uexp}
\end{eqnarray}
generates the coupled $n$-component CH(n,2) system for $p=1,2,\dots,n$, 
\begin{eqnarray} 
\label{CH-n2}
 \partial_t q_p + ( \partial_x q_p  + q_p \partial_x)u_0
 +
( \partial_x q_p  + q_p \partial_x)u_1 -\frac{1}{2}q_{p+2,x} =0
,\end{eqnarray} 
with $q_{n+1}=q_{n+2}=0$.
The differential relations in (\ref{L6}) give $q_1$ and $q_2$ in terms of $u_0$ and $u_1$
in (\ref{L7}) and (\ref{L8}), below. The others $(q_3,\ldots, q_n)$ are independent.

\subsection{CH(n,k) system }   \label{ch(n,k)-sec}

The pattern continues for CH(n,k) upon including  more velocities, $u_0,u_1,\dots,u_{k-1}$, with $u_{k}=-\frac{1}{2}$. 
The $p$-th equation in the integrable $n$-component system CH(n,k) is obtained from  the Lax pair (\ref{L1}) - (\ref{L2}) as 
\begin{eqnarray} 
\label{CH-nk}
 \partial_t q_p
 + 
 \sum_{j=0}^{k-1} 
 (\partial_x q_{p+j} + q_{p+j}\partial_x) u_{j}  
 -\frac{1}{2}q_{p+k,x}=0,
 \\
 \hbox{ with } 
q_{n+1}=q_{n+2}=\dots=q_{n+k}=0.
\nonumber
\end{eqnarray} 
The solutions for $u_0$,...,   $u_{k-1}$ are updated at each time step from $q_1$,...,$q_k$ by imposing the $k+1$ differential relations in (\ref{L6}). 

In the remainder of the paper we will concentrate most of our attention on the two-component systems, including CH(2,1), referred to simply as CH2, as well as CH(2,2) and CH(2,-1). (In the CH(n,k) notation, \emph{negative} values of $k$ refer to positive powers of $\lambda$ in the expansion of $U$ in (\ref{Uexp}).) However,  in the case ${n}=3$ the examples CH(3,1) and CH(3,2) will reveal the graded structure of the Hamiltonian formulations of all of the equations in the CH(n,k) family.  We will also show a few numerical solutions of these equations that provide crucial insight into their pulse-like evolutionary behavior, inviting further investigation.

\section{Examples of two-component CH systems}
\label{examples CH2hierarchy}

\subsection{Example: CH2, or CH(2,1) for $n=2$, $k=1$}

We denote $u_0\equiv u$, $q_1\equiv q$ and $q_2 \equiv \pm \rho^2$, and choose $u_1=-1/2$. 
In this notation, the CH(2,1) system can be written in the form 
\begin{eqnarray}  
q_t\!\!&+&\!\!(\partial_x q + q\partial_x)u \mp \rho \rho_x=0,\label{LL}
\label{CH2-q} \\
\rho_t\!\!&+&\!\!(u\rho)_x=0,\label{LQ}
\label{CH2-rho}
\end{eqnarray} 
where $q=u-u_{xx}+\omega $ and $\omega$ is an arbitrary
constant. 

This is generally known as the CH2 system, and it has been studied extensively. 
The last term in the motion equation in CH2 has a choice of sign ($\mp$). For the positive choice, the CH2 equation may be regarded as a model  of shallow water waves \cite{CI08,HLT09}. Its generalization to higher dimensions is immediate.

\paragraph{Spectral problem for CH2.} The {\bfi spectral problem for CH2} is, as in equations (\ref{CH21-LPairQ}) - (\ref{CH21-LPairU}),
\begin{eqnarray}  \label{SPCH2} \Psi_{xx}&=&\Big(-\lambda^2 \rho^2(x)+\lambda
q(x)+\frac{1}{4}\Big)\Psi. \end{eqnarray}  
This spectral problem is a type of Schr\"odinger equation with an `energy dependent' potential, i.e. it is quadratic in terms of the spectral parameter and moreover the potential functions multiply the spectral parameter
(the so-called weighted problem). There are some common features with
Sturm-Liouville spectral problems, see for example \cite{JJ72,K75,SS97}.
An `energy dependent' spectral problem also appears in the inverse scattering transform 
of an integrable generalization of the Bousinesq equation (Kaup-Bousinesq equation)
\cite{K75}.

\paragraph{Brief history of the CH2 equation.}
The system (\ref{CH2-q})-(\ref{CH2-rho}) representing a two-component generalization of the CH equation
was initially introduced in \cite{SA} as a tri-Hamiltonian system. It was studied further by others, see, e.g., \cite{LZ05,CLZ05,F06,CI08,HLT09,GL10}. 
The known applications of the CH2 model are the following.
\begin{itemize}
\item
In the context of shallow water theory, $u$ can be interpreted as the horizontal
fluid velocity and $\rho$ is related to the water elevation  in the first approximation \cite{CI08,I09}. 
\item
In Vlasov plasma models, CH2 describes the closure of the kinetic moments of the single-particle probability distribution for geodesic motion on the symplectomorphisms \cite{Tr2008,HT09,HoTr2009}. 
\item
In the large-deformation diffeomorphic approach to image matching, the CH2 equation is summoned in a type of matching procedure called metamorphosis \cite{HoTrYo2009}. 
\end{itemize}
The same CH2 system appears as a member of the hierarchy of hydrodynamic chains studied in \cite{SMA}. Its analytical properties such as well-posedness and wave breaking were studied in \cite{ELY07,H09,ZY10,GL10,CL10} and others.

Geometrically, the original CH equation may be interpreted as governing geodesic motion for
an $H^1$ metric that is invariant under the Virasoro group, as found in \cite{M98}. Because of this property, CH is geometrically reminiscent of the Euler rigid body equations, which describe geodesic motion on the rotation group with respect to the metric supplied by the moment of inertia.  The geometric interpretation of CH2 is similar: CH2 is the equation for geodesic motion on the semidirect-product Lie group of diffeomorphisms acting on densities, with respect to the $H^1$ on the horizontal velocity and the $L^2$ norm on the elevation. This is analogous to the finite-dimensional case of an ellipsoidal underwater vehicle (UWV), whose motion may be modelled as geodesics on the Euclidean group of rigid body rotations and translations. This finite-dimensional model of the UWV has the same Lie-Posson bracket as for the Hamiltonian description of the heavy top, so CH2 may also be interpreted analogously to the heavy-top equations.   For additional discussions of geometric aspects of the CH2 system we refer to \cite{HoScSt2009,K07,HoTrYo2009}.

In general, one can show that small initial data of the CH2 system develop into global solutions, while for some initial data wave breaking occurs \cite{ELY07,CI08,H09,ZY10,GL10,CL10}. It is interesting that only the plus sign  ($+$) in (\ref{LL}) corresponds to a positively
defined Hamiltonian and straightforward physical applications to shallow water waves. It would be
interesting to know the physical interpretation of the model with the choice of the minus
sign in (\ref{LL}), since this case is also integrable.

\begin{figure}[ht]
\begin{center}
\includegraphics*[width=0.475\textwidth]{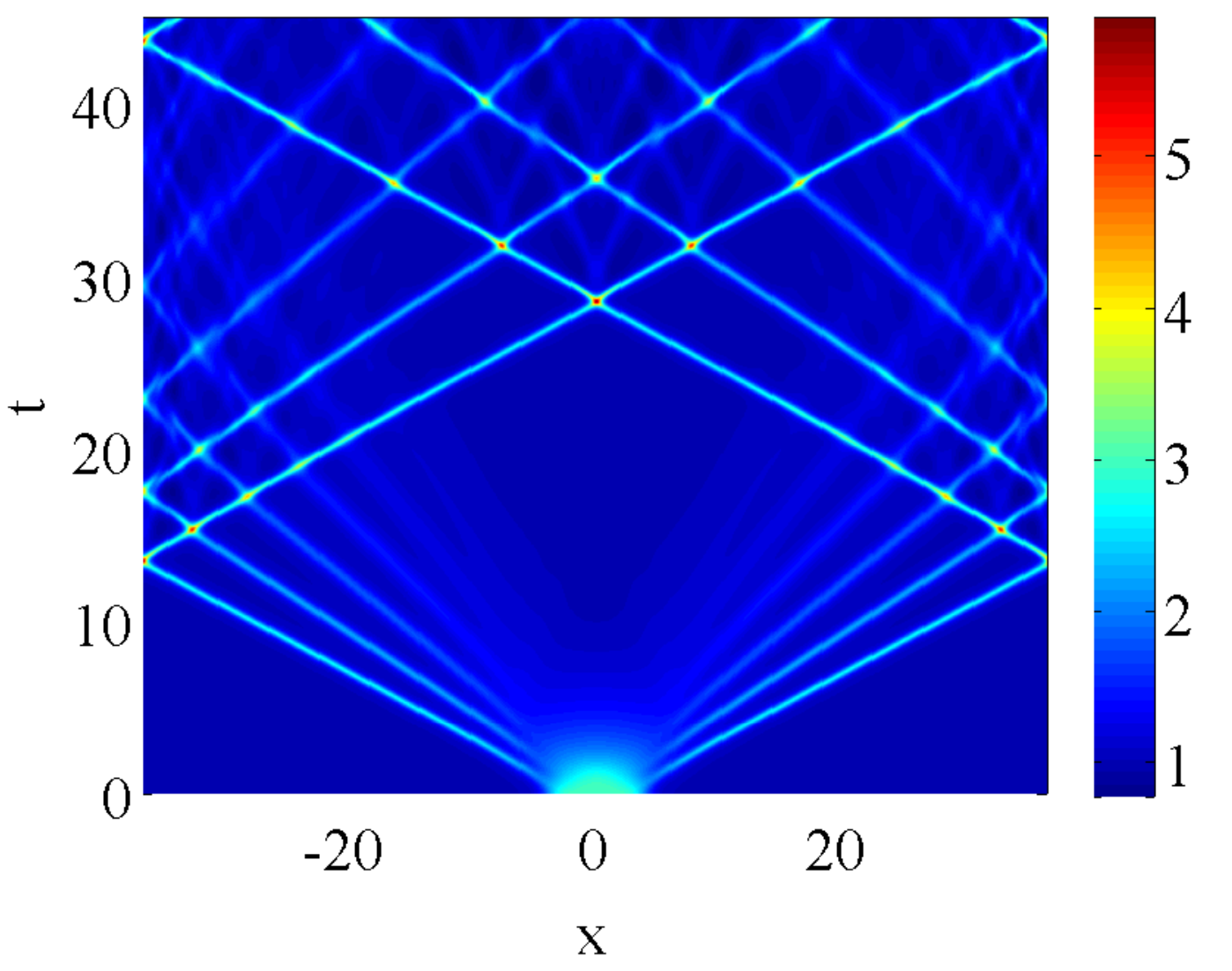}
\includegraphics*[width=0.475\textwidth]{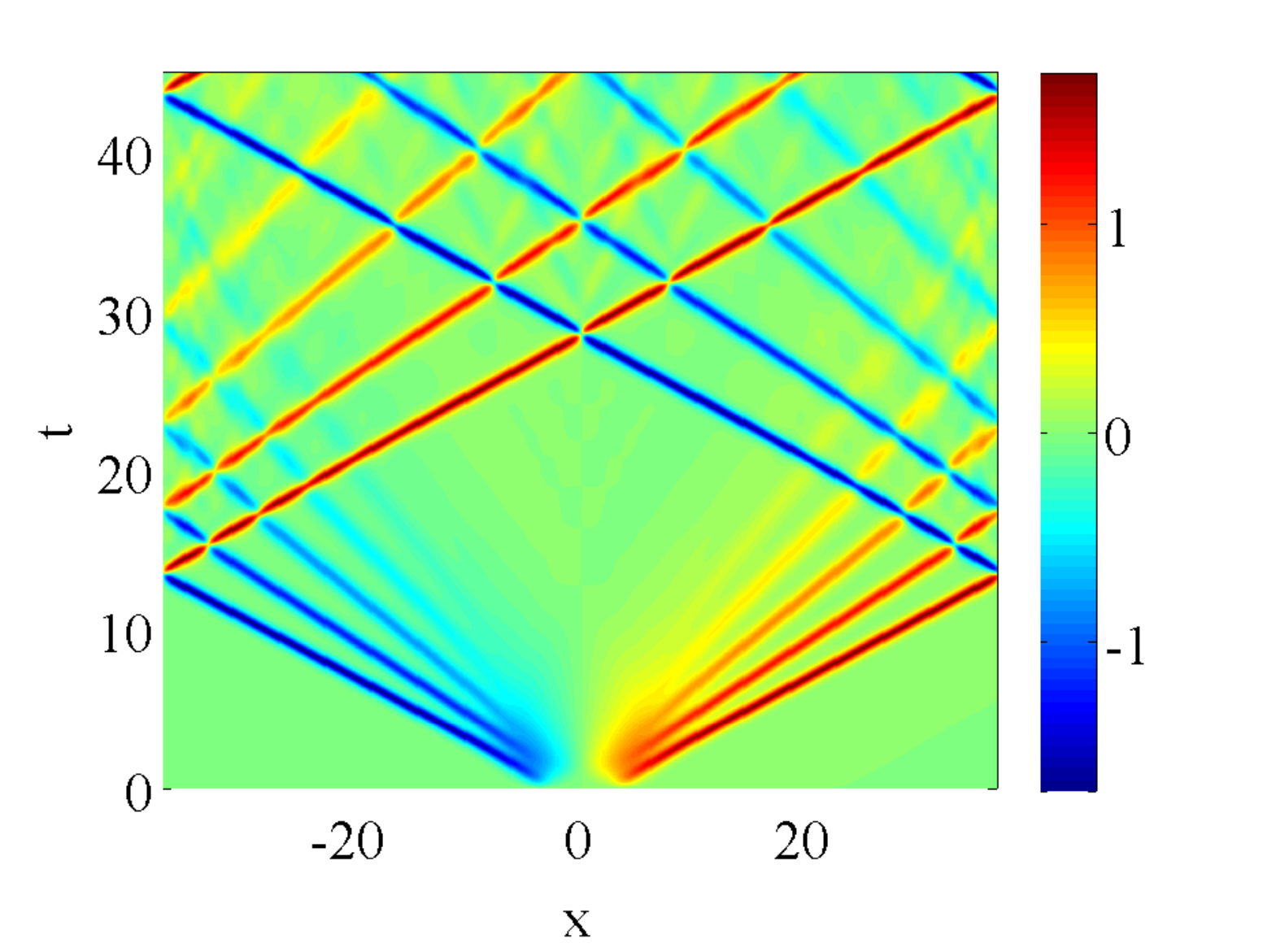}
\end{center}\vspace{-8mm}
\caption{\label{CH2-figs} Dam-break results for the CH(2,1) system in equations (\ref{CH2-q}-\ref{CH2-rho}) show evolution of the density $\rho$ (left panel) and velocity $u$ (right panel), arising from initial conditions (\ref{dambreak-ic}) in a periodic domain. The color bars show positive density on the left and both positive and negative velocity on the right. The soliton solutions are seen to emerge symmetrically leftward and rightward after a finite time, and the evolution of both variables generates more and more solitons propagating in both directions as time progresses.   Figures are courtesy of L. \'O N\'araigh. } 
\end{figure}

\paragraph{Solutions of CH2 for dam-break initial conditions.}
Figure \ref{CH2-figs} plots the evolution of CH2 solutions for $(u,\rho)$ governed by equations (\ref{CH2-q}-\ref{CH2-rho}) with the $+$ sign choice in the periodic domain $\left[-L,L\right]$ with {\bfi dam-break initial conditions} given by
\begin{equation}
u\left(x,0\right)=0,\qquad{\rho}\left(x,0\right)
= 1 +  \tanh(x+a)-\tanh(x-a) 
\,,
\label{dambreak-ic}
\end{equation}
where $a\ll L$.

The dam-break involves a body of water of uniform depth, 
retained behind a barrier, in this case at $x=\pm a$.  If this barrier is
suddenly removed at $t=0$, then the water would flow downward and outward under gravity.  The problem
is to find the subsequent flow and determine the shape of the
free surface.  This question is addressed in the context of shallow-water
theory, e.g., by Acheson \cite{Ach1990}, and thus serves as a typical hydrodynamic
problem of relevance for CH2 solutions with the $+$ sign choice in (\ref{CH2-q}).

The CH2 system invites further generalizations and applications. For example, its two-time generalization is presented in \S\ref{2timeCHeqns} of this paper. 

\subsection{Example: CH(2,2) for $n=k=2$}
CH(2,2) designates the case of two momentum densities $q_1$ and $q_2$ and two velocities $u_0$ and $u_1$. The choice $u_2=-1/2$ automatically solves one of the relations in (\ref{L6}).  The other two differential relations in (\ref{L6}) can then be integrated spatially in $x$, to find the relationships between the momenta and the velocities, as
\begin{eqnarray}  \label{L7} q_{1}&=&u_1-u_{1,xx}+\omega_1,\\
q_2&=&u_0-u_{0,xx}+3u_1^2-u_{1,x}^2-2u_1u_{1,xx}+4\omega_1
u_1+\omega_2,\label{L8}\end{eqnarray} 

where $\omega_{1}$ and $\omega_{2}$ are constants of integration that depend on boundary conditions. For CH(2,2) the evolutionary system (\ref{L6}) yields equations for $q_1$ and $q_2$ given by  
\begin{eqnarray}   
q_{1,t}\!\!&+&\!\! (\partial_x q_1 + q_1\partial_x)u_0 + (\partial_x q_2 + q_2\partial_x)u_1 =0, \label{L9} \\
q_{2,t}\!\!&+&\!\!(\partial_x q_2 + q_2\partial_x)u_0=0.
\label{L10} \end{eqnarray} 
These equations may be solved by first updating $q_2$, then $q_1$ in (\ref{L9}) and (\ref{L10}), followed by inverting the Helmholtz operator twice, first for $u_1$ in (\ref{L7}) and then for $u_0$ in (\ref{L8}). 


\paragraph{Dam-break equivalent problem in $q_{2}$ for CH(2,2).} 
Figure \ref{CH(2,2)-figs} shows the evolution of pulses in the velocity variables $u_0$ (left panel) and $u_1$ (right panel). These pulses arise from an initially localized disturbance in $q_{2}$ with a tanh-squared profile, for $\omega_{1} =\omega_{2}  =0$,
\begin{align*}
q_{2}(x,0) & =\left[1+\tanh\left(x+1\right)-\tanh\left(x-1\right)\right]^{2},\\
u_{1}(x,0) & = q_{1}(x,0)  =0.01,
\end{align*}
that interacts with a constant mean flow in the velocity field $u_{1}$.
The constants $\omega_{1},\omega_{2}$ are set to zero. The initial confined pulse $q_{2}(x,0)$ corresponds to a confined pulse in $u_0$ that propagates steadily rightward and generates a structured dipole pulse in $u_1$ that accompanies the $u_0$ pulse, but may oscillate in polarity as it propagates through a background ``fan'' of smaller slower pulses. This is an interesting scenario whose dynamics will be investigated further elsewhere.

\begin{figure}[t]
\begin{center}
\includegraphics*[width=0.475\textwidth]{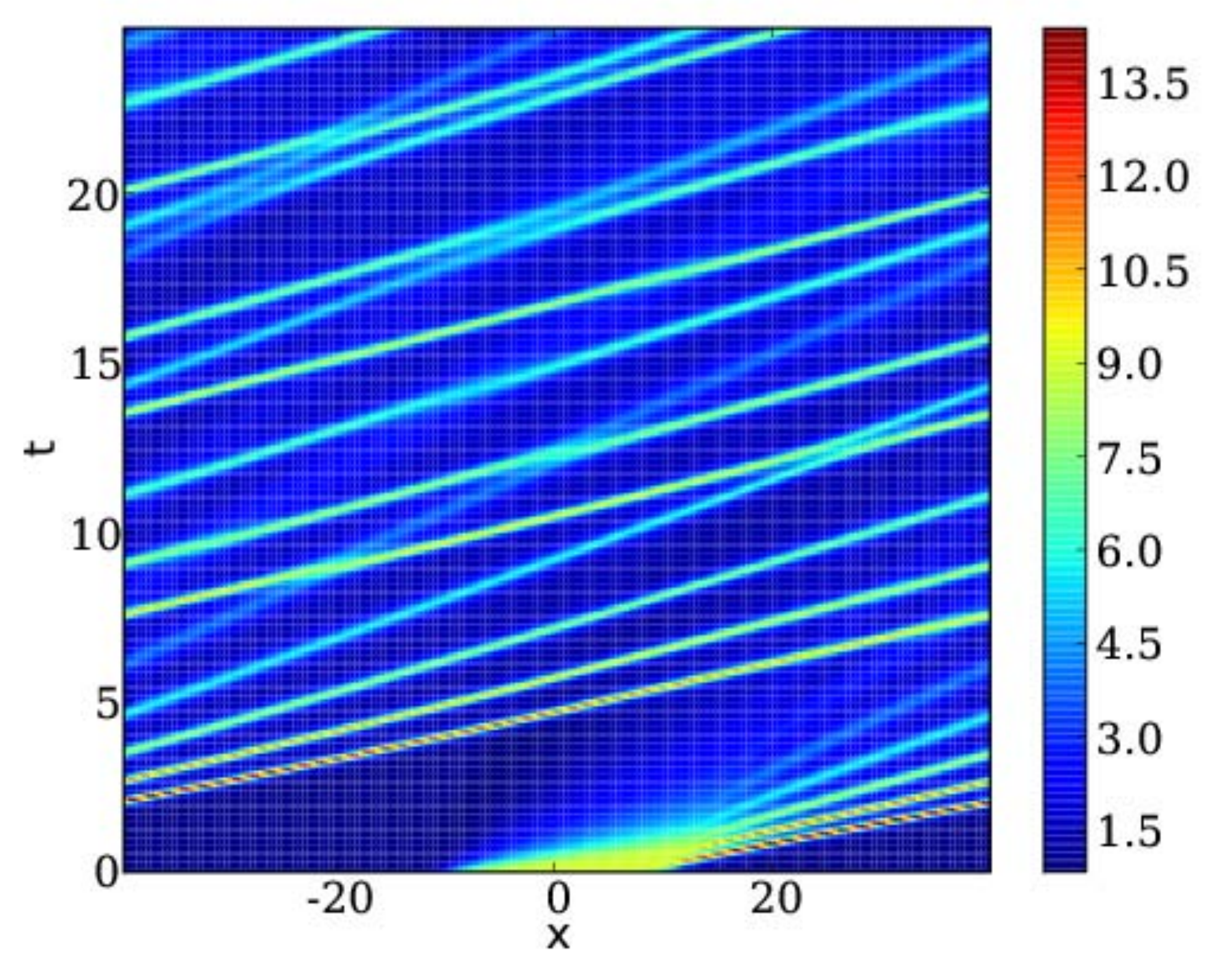}
\includegraphics*[width=0.475\textwidth]{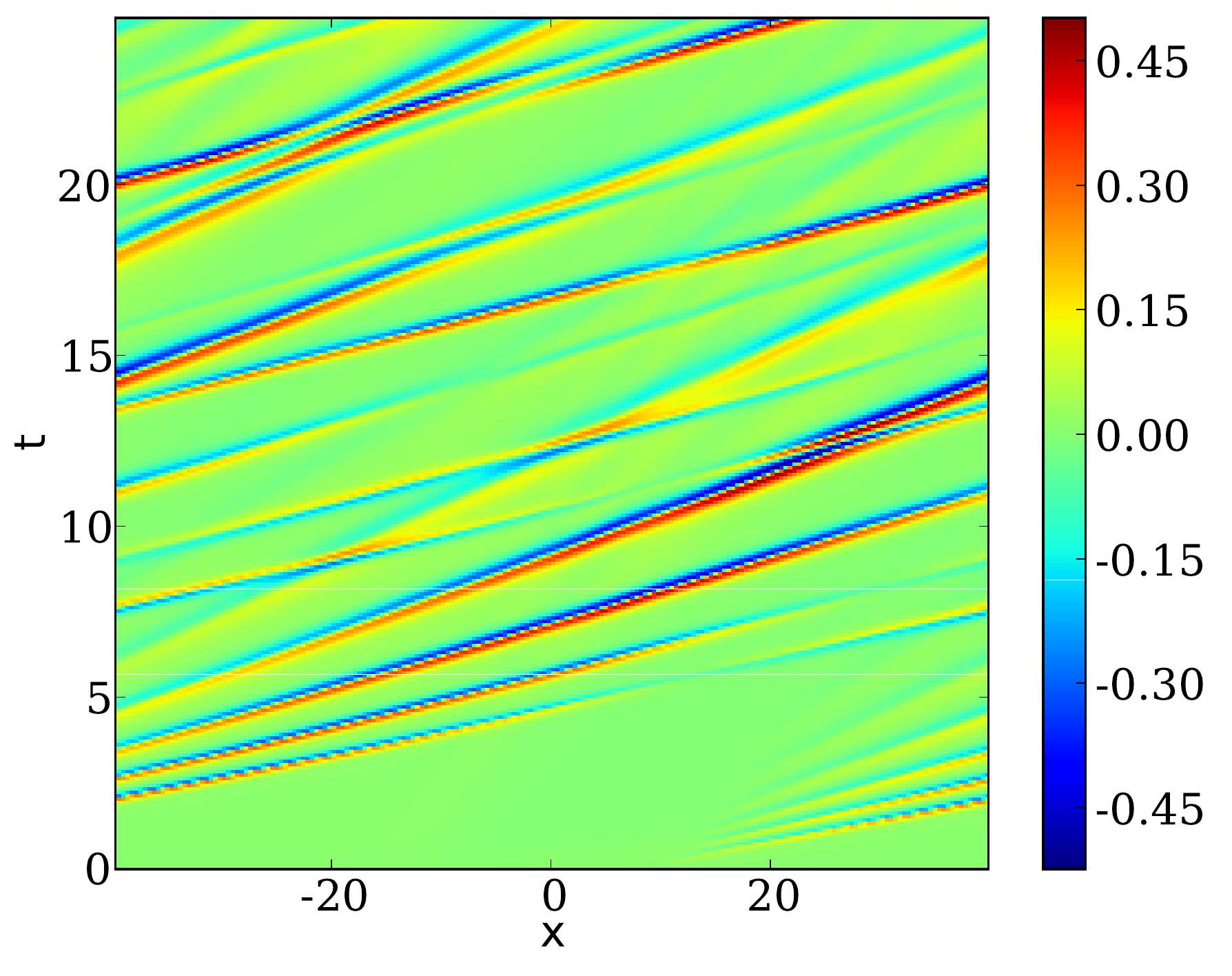}
\end{center}\vspace{-8mm}
\caption{\label{CH(2,2)-figs} This figure shows the dam-break solution behavior of the CH(2,2) system (\ref{L9})-(\ref{L10}), which should be compared with the corresponding results in Figure \ref{CH2-figs} for the system of CH(2,1) equations in (\ref{CH2-q})-(\ref{CH2-rho}).
The dam-break results for the CH(2,2) system show evolution of the velocity variables $u_0$ (left panel) and $u_1$ (right panel), arising from initial conditions for $\sqrt{q_2}$ similar to (\ref{dambreak-ic}) in a periodic domain. A ``fan'' of sequentially smaller, slower pulses is emitted rightward, after which the leading pulses overtake the slower lagging pulses in the periodic domain and suffer elastic collisions that show a variety of different phase shifts, apparently depending on their relative speeds.  Each confined positive pulse in $u_0$ is accompanied by dipole ($\pm$) excitations in $u_1$, whose polarity may occasionally reverse from ($\pm$, red-blue) to ($\mp$, blue-red) and back. Figures are courtesy of J. R. Percival. } 
\end{figure}

\subsection{A semidirect product interpretation of the example CH(2,2)}\label{geom(2,2)}
The system (\ref{L9}-\ref{L10}) for CH(2,2) may be rewritten as 
\begin{eqnarray} 
(\partial_t +\mathcal{L}_{u_0}) (q_2\,dx^2) = 0
,\qquad
(\partial_t +\mathcal{L}_{u_0}) (q_1\,dx^2) + \mathcal{L}_{u_1}\, (q_2\,dx^2)  = 0
\label{22-eq}
,\end{eqnarray} 
where $\mathcal{L}_u$ denotes Lie derivative with respect to the vector field $u$. 

To understand system (\ref{22-eq}) better geometrically, we shall rederive it from Hamilton's principle with a Lagrangian defined on the semidirect product Lie algebra of vector fields  $l(u_0,u_1):\, \mathfrak{X}_0(\mathbb{R})\circledS \mathfrak{X}_1(\mathbb{R})\to \mathbb{R}$. Here $u_0\in\mathfrak{X}_0(\mathbb{R})$ and $u_1\in\mathfrak{X}_1(\mathbb{R})$ are right-invariant smooth vector fields on the real line $\mathbb{R}$. Vector fields in $\mathfrak{X}_0(\mathbb{R})$ act on themselves and on $\mathfrak{X}_1(\mathbb{R})$ by vector cross product, while the action of $\mathfrak{X}_1(\mathbb{R})$ on itself is assumed to be by simple vector addition. That is, vector fields in $\mathfrak{X}_1(\mathbb{R})$ are {\bfi advected quantities}, see, e.g., \cite{HoMaRa1998}.

\paragraph{Definitions: Semi-direct product.}
The semidirect product Lie algebra action $\mathfrak{X}_0(\mathbb{R})\circledS \mathfrak{X}_1(\mathbb{R})$ is defined by
\begin{equation}
[(X_0,X_1), (Y_0,Y_1) ]
=
( [X_0,Y_0], [X_0,Y_1] + [X_1,Y_0]),
  \label{sdpLieAction}
\end{equation}
where $[\,\cdot\,,\,\cdot\,]$ is the commutator of vector fields in natural notation.
This commutator defines the adjoint action
\begin{equation}
{\rm ad}_{(X_0,X_1)}(Y_0,Y_1) = - [(X_0,X_1), (Y_0,Y_1) ]
= ({\rm ad}_{X_0}Y_0,\, {\rm ad}_{X_0}Y_1 + {\rm ad}_{X_1}Y_0 )
  \label{sdp-adAction}
\end{equation}
and the $L^2$ pairing with $(\alpha, \beta)\in \mathfrak{X}_0^*\circledS \mathfrak{X}_1^*$ yields the coadjoint action
\begin{eqnarray} 
&& \left\langle {\rm ad}^*_{(X_0,X_1)}(\alpha, \beta),\,
(Y_0,Y_1) \right\rangle 
=\left\langle (\alpha, \beta),\,
{\rm ad}_{(X_0,X_1)}(Y_0,Y_1) \right\rangle 
\nonumber\\
&&
=\left\langle (\alpha, \beta),\,
({\rm ad}_{X_0}Y_0,\, {\rm ad}_{X_0}Y_1 + {\rm ad}_{X_1}Y_0 )\right\rangle 
\nonumber \\
&&
=\left\langle \alpha ,\, {\rm ad}_{X_0}Y_0 \right\rangle 
+
\left\langle   \beta ,\,  {\rm ad}_{X_0}Y_1 + {\rm ad}_{X_1}Y_0 \right\rangle 
  \label{sdp-ad*Action}\\
&&
=\left\langle {\rm ad}^*_{X_0}\alpha + {\rm ad}_{X_1}^* \beta,\, Y_0 \right\rangle 
+
\left\langle  {\rm ad}_{X_0}^* \beta ,\,  Y_1 \right\rangle 
\nonumber\\
&&
=\left\langle ({\rm ad}^*_{X_0}\alpha + {\rm ad}_{X_1}^* \beta, {\rm ad}_{X_0}^* \beta) 
,\, (Y_0, Y_1) \right\rangle 
.\nonumber
\end{eqnarray} 

See \cite{BrGaHoRa2010} for more background and derivations of the formulas for the adjoint and coadjoint actions of the semidirect product  Lie group ${\rm Diff}_0(\mathbb{R})\circledS{\rm Diff}_1(\mathbb{R})$ and its Lie algebra of right-invariant vector fields $\mathfrak{X}_0\circledS \mathfrak{X}_1$.

We shall rederive system (\ref{22-eq}) from Hamilton's principle by using the Euler-Poincar\'e theory, as reviewed for continuum mechanics, e.g., in \cite{HoMaRa1998}. We then pass to its Hamiltonian formulation in terms of a Lie-Poisson bracket by performing a Legendre transformation.
In the Euler-Poincar\'e framework, we have the following.

\begin{theorem}[Euler-Poincar\'e formulation of CH(2,2)]$\,$\\
Hamilton's principle $\delta S=0$ with $S=\int l(u_0,u_1)\,dt$ yields the CH(2,2) system in equations (\ref{22-eq}) for the Lagrangian
\begin{eqnarray} 
l(u_0,u_1) = 
\int_\mathbb{R} u_0u_1 + u_{0,x}u_{1,x} + \omega_1u_0 + \omega_2u_1
 + 2u_1\left(u_1^2 + u_{1,x}^2\right)\,dx
\label{CH22-Lag}
,\end{eqnarray} 
for constrained variations of $u_0$ and $u_1$ of the semidirect product form in (\ref{sdpLieAction}),
consisting of
\begin{eqnarray} 
\delta u_0 = \partial_t \xi_0 + [ u_0, \xi_0 ] = \partial_t \xi_0 - {\rm ad}_{u_0} \xi_0  
,\\
\delta u_1 = \partial_t \xi_1 + [ u_0, \xi_1 ] + [ u_1, \xi_0 ]  
= \partial_t \xi_1 - {\rm ad}_{u_0} \xi_1 - {\rm ad}_{u_1} \xi_0
\label{CH22-var}
.\end{eqnarray} 

\end{theorem}

\begin{proof}
By direct calculation, Hamilton's principle with the Lagrangian (\ref{CH22-Lag}) implies
\begin{eqnarray} 
0 = \delta S &=& \int \langle q_1,\,\delta u_0 \rangle + \langle q_2,\,\delta u_1 \rangle \,dt
\nonumber\\
&=& \int \langle q_1,\,\partial_t \xi_0 - {\rm ad}_{u_0} \xi_0 \rangle 
+ \langle q_2,\, \partial_t \xi_1 - {\rm ad}_{u_0} \xi_1 - {\rm ad}_{u_1} \xi_0 \rangle \,dt
\nonumber\\
&=& - \int \langle \partial_t  q_1 + {\rm ad}^*_{u_0}q_1 + {\rm ad}^*_{u_1}q_2,\, \xi_0 \rangle 
+ \langle \partial_t  q_2 + {\rm ad}^*_{u_0}q_2,\,  \xi_1  \rangle \,dt
\,,\nonumber
\end{eqnarray} 
where $q_1$ and $q_2$ are given in (\ref{L7}) and (\ref{L8}), respectively, and we have used the semidirect product Lie algebra action defined in (\ref{sdpLieAction}). The coadjoint operation ${\rm ad}^*$ is defined using the $L^2$ pairing as in, e.g.,
\begin{eqnarray} 
\langle q_2,\,{\rm ad}_{u_0} \xi_1 \rangle 
=
\langle {\rm ad}^*_{u_0} q_2,\, \xi_1 \rangle 
=
\langle \mathcal{L}_{u_0} q_2,\, \xi_1 \rangle 
=
\langle (\partial_x q_2 + q_2\partial_x)u_0,\, \xi_1 \rangle 
,\end{eqnarray} 
which is a repeated pattern in the system (\ref{22-eq}). 

\end{proof}
%

\paragraph{Legendre transform to the CH(2,2) Hamiltonian formulation.}
The Legendre transform for CH(2,2) is given by
\begin{eqnarray} 
h(q_1,q_2) = \langle (q_1,q_2),\,(u_0,u_1)\rangle - l(u_1,u_2)
\,,\quad
\frac{\delta h}{\delta m}=u
\,,\quad
\frac{\delta h}{\delta u}=0 = m - \frac{\delta l}{\delta u}
\,,
\end{eqnarray} 
where $\langle \,\cdot\,,\,\cdot\,\rangle$ denotes $L^2$ pairing on the real line. The corresponding variations yield
\begin{eqnarray} 
\delta h =
 \left\langle  (u_0,u_1),\, (\delta q_1,\,\delta q_2) \right\rangle
+ \left\langle  
\left(q_1 - \frac{\delta l}{\delta u_0},\, q_2 - \frac{\delta l}{\delta u_1} \right) 
,\, (\delta u_0,\, \delta u_1)  \right\rangle
.
\end{eqnarray} 
Hence the pairs $(q_1,q_2)$ and $(u_0,u_1)$ are dual variables with respect to the Legendre transform of the CH(2,2) Lagrangian in (\ref{CH22-Lag}). Thus, the differential relations in the Lax pair formulation of CH(2,2) appearing in  (\ref{L9}) and (\ref{L10}) Legendre  transform into dual momenta for the Hamiltonian formulation. 

\paragraph{Lie-Poisson Hamiltonian form of CH(2,2).}
The CH(2,2) system in (\ref{L9}-\ref{L10}) or  (\ref{22-eq}) may now be cast into Lie-Poisson Hamiltonian form, as
\begin{equation}
\begin{bmatrix}
\partial_t q_1 \\
\partial_t q_2
\end{bmatrix}
= - 
\begin{bmatrix}
\partial_x q_1 + q_1 \partial_x & \partial_x q_2 + q_2 \partial_x \\
\partial_x q_2 + q_2 \partial_x &  0
\end{bmatrix}
\begin{bmatrix}
\delta h /\delta q_1 = u_0 \\
\delta h /\delta q_2 = u_1
\end{bmatrix}
  ,
  \label{SDP Ham form}
\end{equation}
The Hamiltonian operator yields the {\bfi Lie-Poisson bracket} defined on the dual to the {\bfi semidirect product} Lie algebra of vector fields $\mathfrak{X}_0(\mathbb{R})\circledS\mathfrak{X}_1(\mathbb{R})$.  The Lie algebra action for the semidirect product is defined by (\ref{sdpLieAction}) and dual coordinates are $q_1\in \mathfrak{X}_0^*(\mathbb{R})$ and $q_2\in \mathfrak{X}_1^*(\mathbb{R})$. 

\begin{remark}\rm
The Lie-Poisson Hamiltonian form (\ref{SDP Ham form})
\[
\partial_t (q_2\,dx^2) =-\mathcal{L}_{u_0}(q_2\,dx^2)
\]
may be interpreted as saying that the 1-form density $(q_2dx^2)$ evolves in time $t$ by the action of ${\rm Diff}_0(\mathbb{R})$ on its initial conditions. That is,
\[
\frac{d}{dt}(q_2\,dx^2) = 0
\quad\hbox{along}\quad
\frac{dx}{dt} = u_0 (t, x(t))
\,.
\]
Quantities that evolve this way to remain invariant along the characteristic paths of a flow velocity  in ideal fluid mechanics are said to be {\bfi advected}, or {\bfi frozen into the flow}. Equation (\ref{22-eq}) shows that the process is not passive, though; because the dynamics of $q_1$ is affected by a force depending on $q_2$ and the corresponding velocities that are obtained from the differential relations  in (\ref{L9}) and (\ref{L10}).

The other Hamiltonian structure for this example would be interesting to know. However, knowing it is not necessary for the sake of generating its integrable hierarchy, because we already have its isospectral problem and Lax pair.

\end{remark}

\paragraph{Analogy with rotating tops.}
If the CH(2,2) problem specified here for a right-invariant Lagrangian on $\mathfrak{X}\circledS \mathfrak{X}$ had been expressed instead on $\mathfrak{so}(3) \circledS \mathfrak{so}(3)$ for a \emph{left-invariant} Lagrangian, the result would have been interpreted as the dynamics of a rotating top in a potential force field, as discussed in \cite{Bog1985}. The dynamics in this case is expressible in Hamiltonian form as
\begin{equation}
\begin{bmatrix}
\mathbf{\dot{q}}_1 \\
\mathbf{\dot{q}}_2
\end{bmatrix}
= 
\begin{bmatrix}
\mathbf{q}_1\times & \mathbf{q}_2\times \\
\mathbf{q}_2\times &  0
\end{bmatrix}
\begin{bmatrix}
\delta h /\delta \mathbf{q}_1 = \mathbf{u}_0 \\
\delta h /\delta \mathbf{q}_2 = \mathbf{u}_1
\end{bmatrix}
  \,,
 \label{top-prob}
\end{equation}
for angular momenta $(\mathbf{q}_1,\mathbf{q}_2)\in\mathbb{R}^3\times\mathbb{R}^3$
and their corresponding angular velocities $(\mathbf{u}_0,\mathbf{u}_1)\in\mathbb{R}^3\times\mathbb{R}^3$ and Hamiltonian $h=\frac12( \mathbf{q}_1\cdot\mathbf{u}_0 +  \mathbf{q}_2\cdot\mathbf{u}_1)$.
This Hamiltonian matrix defines a Lie-Poisson bracket on the dual of the semidirect-product Lie algebra $\mathbb{R}^3\circledS\mathbb{R}^3_0$ in which the first $\mathbb{R}^3$ acts on itself and on the second, $\mathbb{R}^3_0$, by vector cross product, while the action of the second $\mathbb{R}^3_0$ on itself is by simple vector addition. 

\section{Other examples of two-component {CH} generalizations}\label{eqns-hierarchy-sec}

In this section we discuss two other examples of equations in the integrable CH2 hierarchy. These are: (i) the equation denoted CH(2,-1) in the CH2 hierachy that occupies the ``Dym position'' in the KdV hierachy; and (ii) the CH(2,1) equations with two time variables.  

\subsection{The CH2 Dym equation or CH(2,\,-1)}\label{CH2Dym}

\begin{figure}[h]
\begin{center}
\includegraphics*[width=0.50\textwidth]{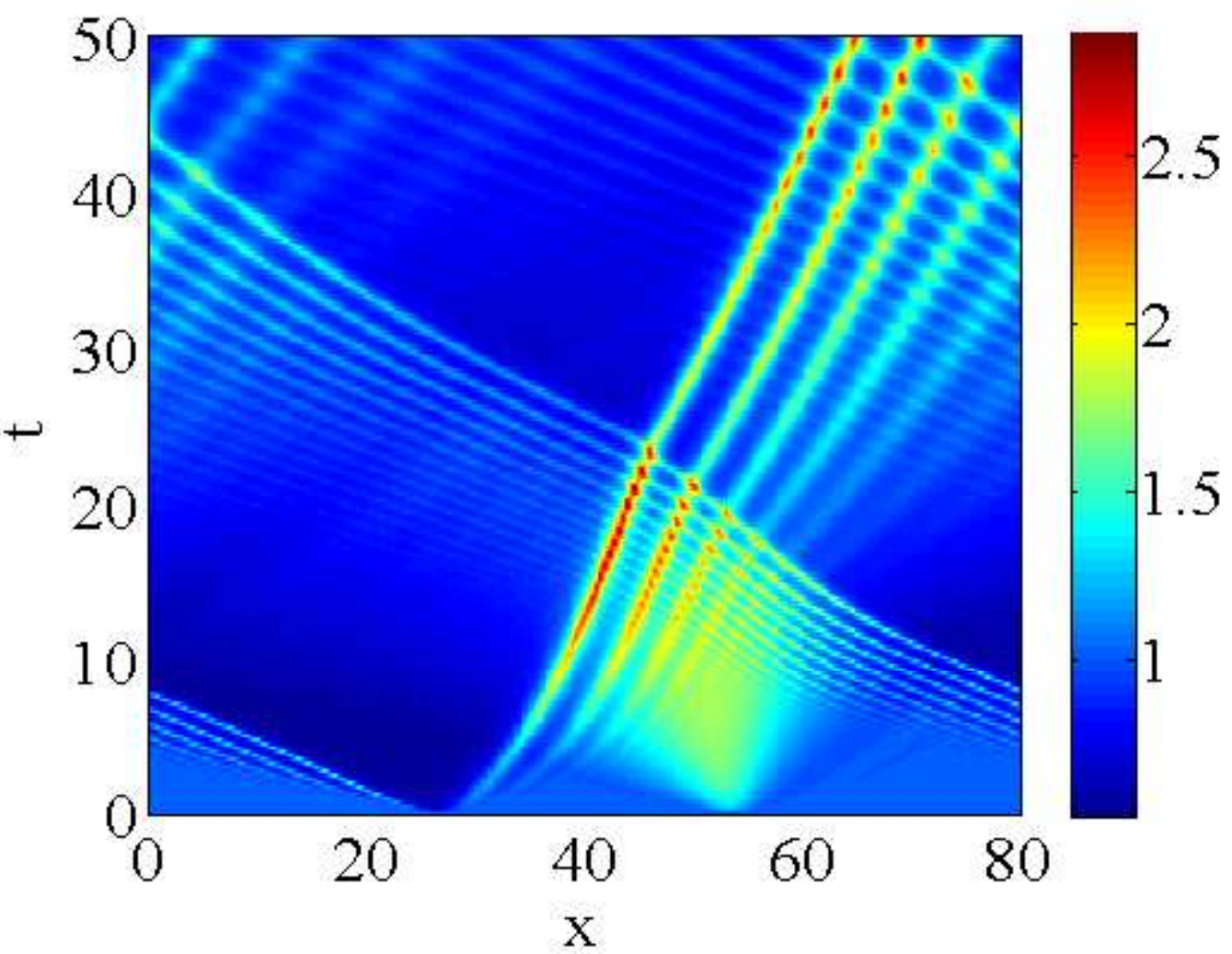}
\includegraphics*[width=0.48\textwidth]{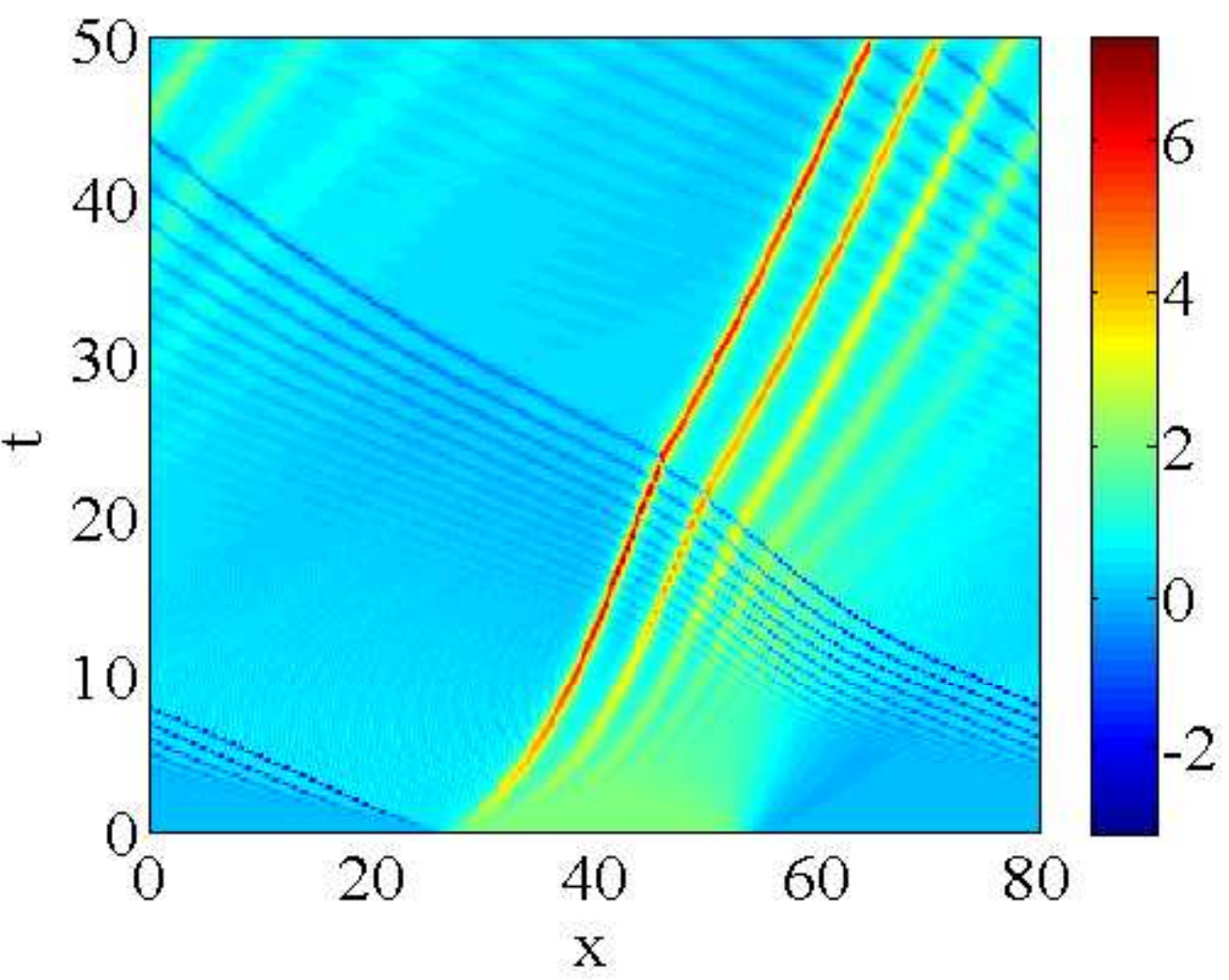}
\end{center}\vspace{-5mm}
\caption{
Results are shown for the evolution of the CH(2,-1) system (\ref{rho-dot}-\ref{q-dot}) with $\epsilon=-1$ for $\rho$ (left) and $q$ (right), arising in a periodic domain of length $L=80$ from initial conditions that represent a dam-break $q(x,0)=\tanh((x-L_1)/\alpha)-\tanh((x-L_2)/\alpha),\quad \rho(x,0)=1$ with $\alpha=1$, $L_1=L/3$, $L_2=2 L/3$. Soliton solutions are seen to emerge and propagate in both directions. The head-on collision process produces a slight refraction of the soliton trajectories, unlike the CH(2,1) case.  
Figures are courtesy of V. Putkaradze.
\label{DymCH2-fig} }
\end{figure}

The CH(2,2) system (\ref{L7}) -- (\ref{L10}) sits in a hierarchy of integrable equations that share the same spectral problem (\ref{SPCH2}), as well as the main representative CH(2,1). There are other members of this hierarchy, for which $U(x,\lambda)$ contains \emph{positive} powers of $\lambda$. A simple interesting example of this kind has 
\begin{eqnarray} 
Q=-\,\lambda^2 \rho^2+\lambda q + 1/4
\quad\hbox{and}\quad
U
=
\lambda u
\,,
\label{CH2-1}
\end{eqnarray} 
Extending our notation, we can denote the resulting system as CH(2,-1).%
\footnote{ It is possible to extend the expansions in powers of $\lambda$ in (\ref{L3}), (\ref{L4}) in both directions for both $Q$ and $U$. The equation chosen for analysis here is only a single step in this extension.}
%
Substituting (\ref{CH2-1}) into (\ref{L5}) leads to equations denoted as CH(2,\,-1):
\begin{eqnarray}
\rho_t +\left(\frac{ q}{\rho^2} \right)_x &=& 0
\,,\label{rho-dot}\\
 q_t - \left(\big(1-\partial_x^2 \,\big)\frac{1}{\rho}\right)_x &=& 0
\,.\label{q-dot}
\end{eqnarray}
where we have used $ u \rho =K$, obtained from the differential relation arising in the $\lambda^3$ term, and have set the constant value $K=-2$.  Some of its solution behavior is shown in Figure \ref{DymCH2-fig}

{\bf Remarks.}\\
Here are a few remarks about the CH(2,-1) system in equations (\ref{rho-dot}) and (\ref{q-dot}):
\begin{itemize}
\item
This coupled nonlinear system is at the position in the CH2 hierarchy that corresponds to the modified Dym equation, first introduced as a tri-Hamiltonian system in \cite{SA}. 

\item
The CH(2,-1) equations combine to produce the {\bf nonlinear wave equation} 
\begin{eqnarray}
q_{tt} =  \big(1-\partial_x^2 \,\big) \left(\partial_x \frac{1}{\rho^2}\partial_x \frac{1}{\rho^2}\right)q
\,.\label{qrho-wave}
\end{eqnarray}
Linearizing this equation around $q=0$ and $\rho=1$ yields the dispersion relation for a plane wave $\exp(i(kx-\omega t)$ with wave number $k$ and frequency $\omega$ as 
\[
\omega^2(k)=(1+k^2)k^2
\,.
\]
Accordingly, the phase speed of the linearized plane waves is $\omega/k = \sqrt{1+k^2}$, so the higher wave numbers travel faster. This type of dispersion relation is not unfamiliar: it is the same as for 
time-dependent Euler-Bernoulli theory for an elastic beam with with both bending and vibration response \cite{Re2007}. 

\item 
The CH(2,-1) {\bf travelling wave solutions} $v(\xi)=1/\rho(\xi)$ with $\xi=x-ct$ for $c>0$ conserve the energy $E$ given by 
\begin{eqnarray}
2E = (v')^2 - \left(\frac{c}{v}-I\right)^2 - (v+J)^2
\,,\label{trav-wave}
\end{eqnarray}
with integration constants $I$ and $J$ defined by 
\begin{eqnarray}
I = \frac{c}{v} - q v^2
\quad\hbox{and}\quad
J = v^{''} - v + cq
\,.\label{integ-const}
\end{eqnarray}
For the travelling wave $E=0$, and when $I=0=J$, as well, then the solution is
given  by
\begin{eqnarray}  
c\rho(\xi) = \sqrt{{\rm sech}(2(\xi-\xi_0))\,}, 
\qquad \xi_0=\text{constant and} \qquad
q(\xi)=c\rho^3(\xi)
.
\end{eqnarray} 
This is a confined travelling wave pulse in both $\rho$ and $q$. 

\item 
The coupled CH(2,-1) system (\ref{rho-dot}-\ref{q-dot}) may also be written in {\bf Hamiltonian form}, as
\begin{equation}
\partial_t 
\begin{bmatrix}
\rho \\
\epsilon q
\end{bmatrix}
=
\begin{bmatrix}
\partial_x  & 0 \\
0 & \partial_x-\partial_x^3
\end{bmatrix}
\begin{bmatrix}
-q/\rho^2 = \delta h / \delta \rho 
\\
1/\rho = \delta h / \delta q 
\end{bmatrix}
,\quad\hbox{with}\quad
h := \int (q/\rho)\,dx\,.
\end{equation}
This Hamiltonian operator yields the Poisson bracket dynamics,
\begin{equation}
\frac{dF}{dt}
=
\{F,H\} = \int \left(\frac{\delta F}{\delta \rho} \partial_x \frac{\delta H}{\delta \rho}
+ \frac{\delta F}{\delta q} \left(\partial_x-\partial_x^3\right) \frac{\delta H}{\delta q}\right)dx
\,.
\end{equation}

The energy conservation law may be expressed in conservative form as 
\begin{equation}
\partial_t \left( \frac{q}{\rho} \right)
+ 
\partial_x \left(-\,\frac{1}{2}\left(\frac{q}{\rho^2}\right)^2 
+
\frac{1}{2\rho^2}
-
\frac{1}{\rho}\partial_x^2\frac{1}{\rho}
+
\frac{1}{2}\left(\partial_x\frac{1}{\rho} \right)^2
\right)
=
0
\,.
\end{equation}

\item 
The CH(2,-1) equations (\ref{rho-dot}) and (\ref{q-dot}) may also be {\bf written in $n$ dimensions} as 
\begin{eqnarray}
\rho_t + \nabla\cdot \left(\frac{ \mathbf{q} }{\rho^2} \right) &=& 0
\,,\label{rho-dot-n}\\
 \mathbf{q}_t - \nabla \left(\big(1-\nabla ^2 \,\big)\frac{1}{\rho}\right) &=& 0
\,,\label{q-dot-n}
\end{eqnarray}
for $\rho\in\mathbb{R}$ and $\mathbf{q}\in\mathbb{R}^n$.

\item

The consistency among the CH(n,k) equations can be demonstrated by combining them.
The choice, $Q=\epsilon_1\lambda^2
\rho^2+\lambda q + \frac{1}{4}$ and $U=-\frac{1}{2\lambda}+u+2\epsilon_2
\lambda/\rho$ with $\epsilon_{1,2}=\pm 1$ in (\ref{L5}), for
example, produces another nonlinear integrable system:%
\footnote{
One may compare this with 
$Q=-\,\lambda^2 \rho^2+\lambda q + 1/4
\quad\hbox{and}\quad
U
=
\lambda u
$ for CH(2,-1) in (\ref{CH2-1}).}
\begin{eqnarray}
\epsilon_1\Big(\rho_t +(u\rho)_x \Big)+\epsilon_2\left(\frac{ q}{\rho^2}
\right)_x &=& 0
\,,\label{new rho-dot}\\
\ q_t + (\partial_x q+q\partial_x)u-\epsilon_1 \rho \rho_x
+\epsilon_2\left(\left(1-\partial_x^2 \right)\frac{1}{\rho}\right)_x &=& 0
\,,\label{new q-u-rho syst}
\end{eqnarray}
with differential relation $q_x=u_x - u_{xxx}$.
This system reduces to CH2 for $\epsilon_2=0$.

\end{itemize}

\subsection{Equations in the CH2 hierarchy with two time variables}\label{2timeCHeqns}

There is also an integrable CH2 system with two `time' variables ($t$ and $y$). In particular, consider the system%
\footnote{A two-time version of the CH equation has been considered previously in \cite{Iv2009}.}
The two-time CH2 system with $m=U_x-U_{xxx}$ is
\begin{eqnarray}  
m_t+2 U_{yx} m + (U_y+\gamma) m_x 
+ \rho 
\rho_y &=&0
, \label{m-eqn} \\
\rho_t + \Big((U_y +\gamma )\rho \Big)_x&=&0. 
\label{rho-eqn} 
\end{eqnarray} 

This system can be written equivalently in a hydrodynamic form as 
\begin{eqnarray}  
(m/\rho^2)_t + (U_y +\gamma )(m/\rho^2)_x&=& -\, \rho^{-1}\rho_y
, \label{m-eqn2} \\
\rho_t + \Big((U_y +\gamma )\,\rho \Big)_x&=&0, 
\label{rho-eqn2} 
\end{eqnarray} 
which shows that it has only one characteristic velocity, $dx/dt=(U_y +\gamma)$. 

The two-time CH2 system can also be written as the compatibility condition for the following linear system ({\bfi Lax pair}) with a constant spectral parameter $\zeta$: 
\begin{eqnarray}  
\Psi_{xx}&=&\Big(-\zeta^2\rho^2+\zeta m+\frac{1}{4}\Big)\Psi
, \label{ev-prob}
\\
\Psi_{t}-\frac{1}{2\zeta}\Psi_y&=&-\,(U_y+\gamma)\Psi_x+\frac{1}{2}U_{yx}\Psi
.
\label{evol-y}
\end{eqnarray}

\begin{figure}[ht]
\begin{center}
\includegraphics[width=0.47\textwidth,angle=0]{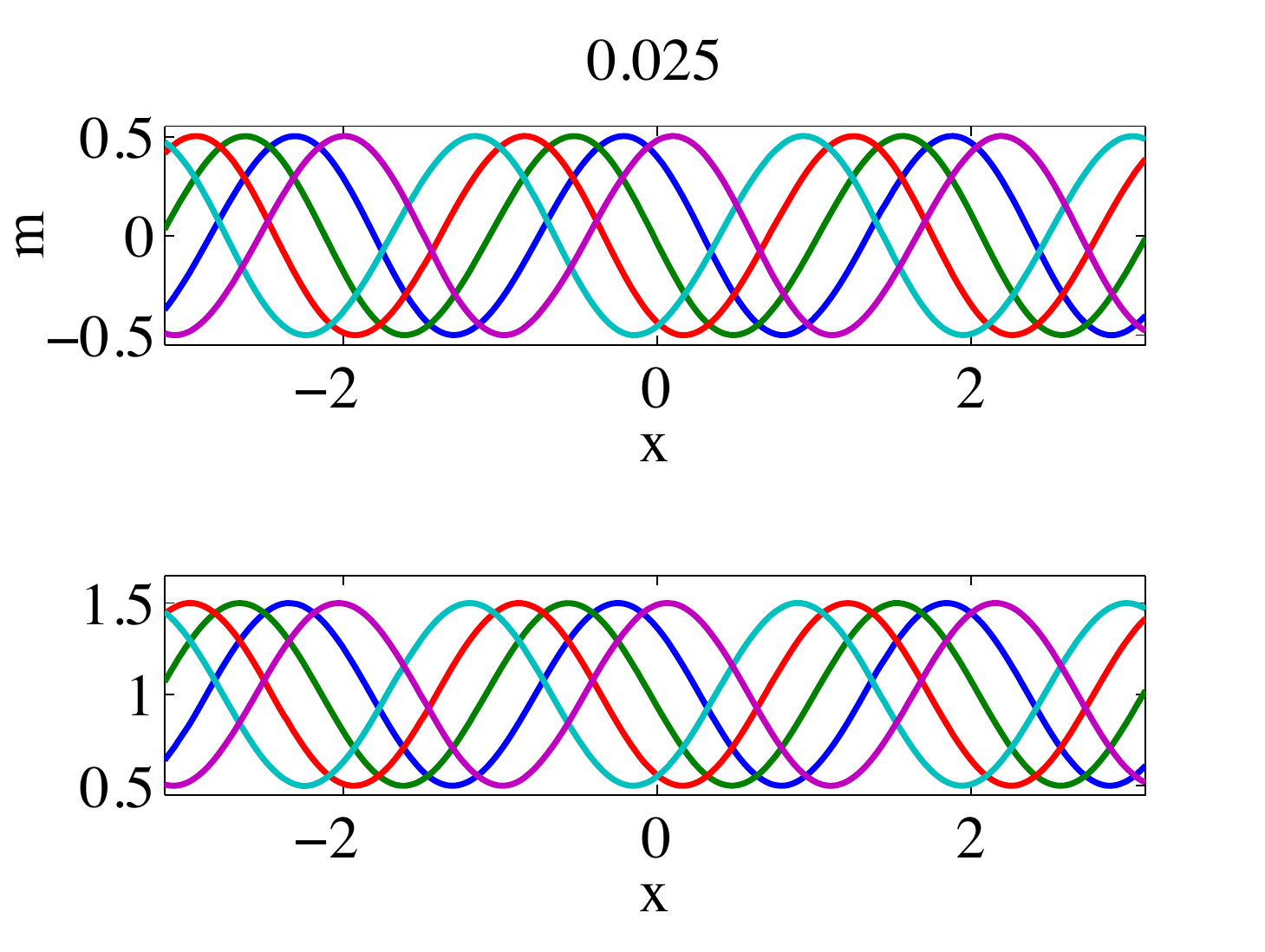}
\includegraphics[width=0.47\textwidth,angle=0]{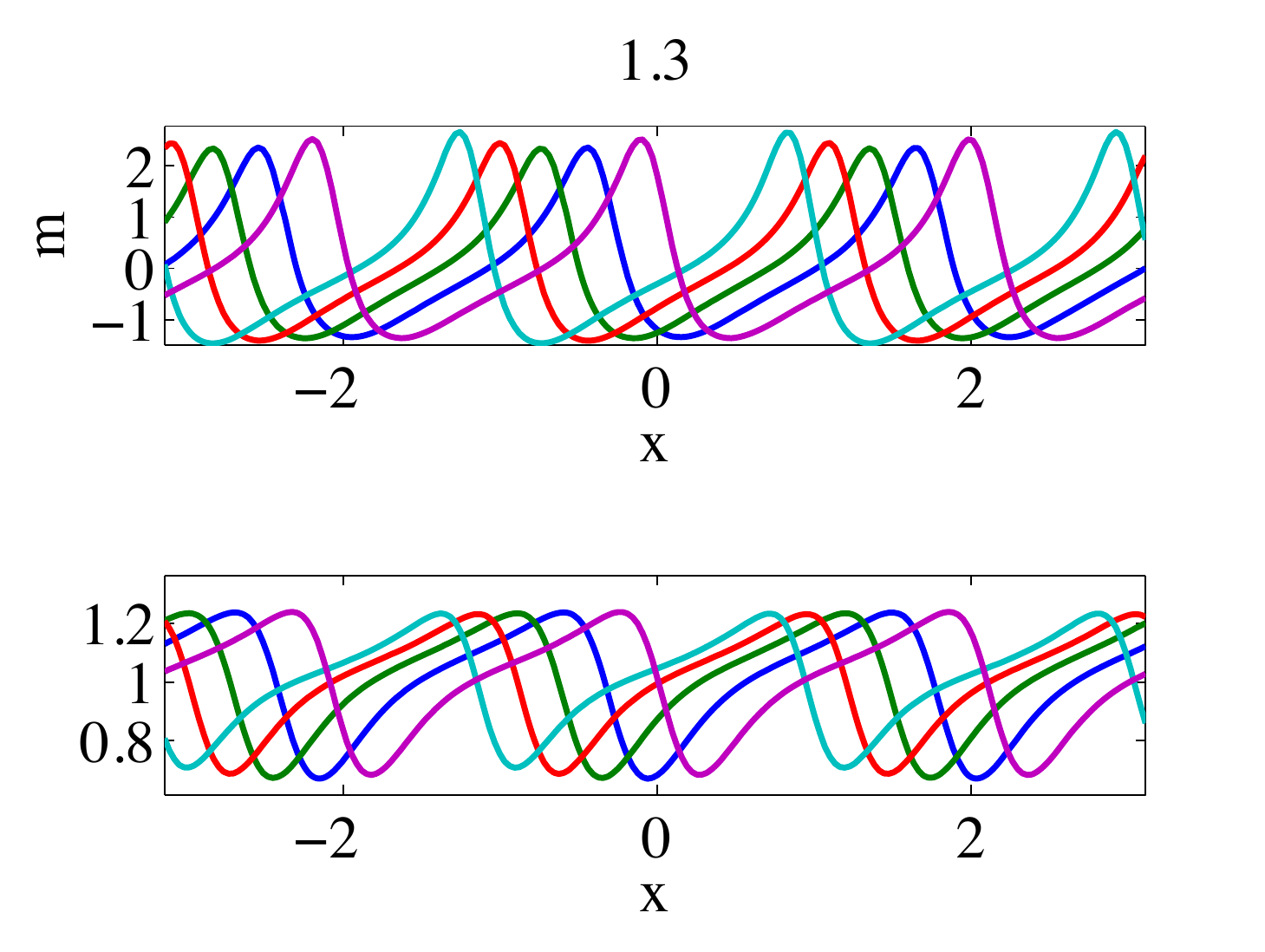}
\includegraphics[width=0.66\textwidth,angle=0]{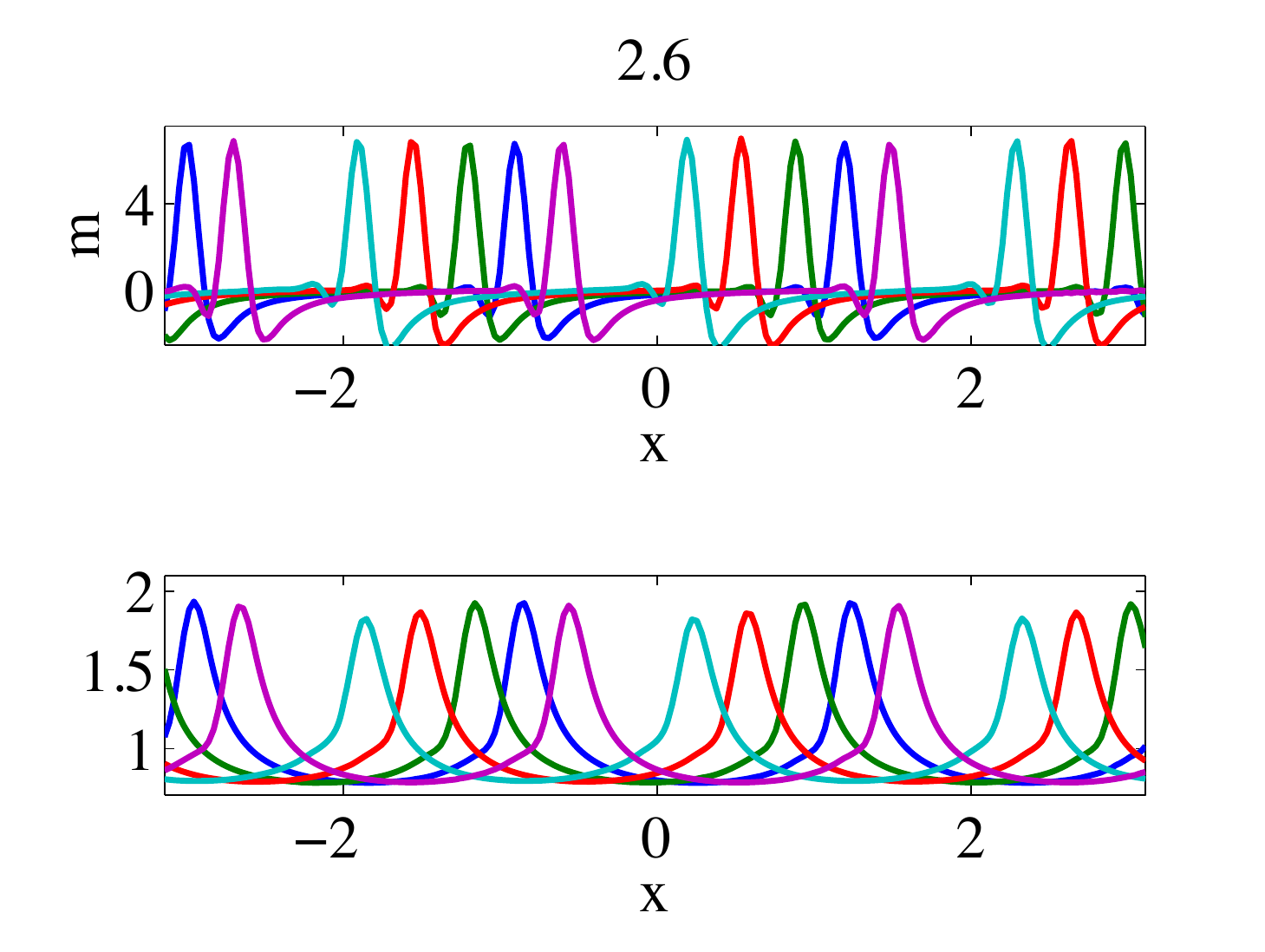}
\end{center}\vspace{-8mm}
\caption{
An initially sinusoidal wave train (plotted with $m$ above and $\rho$ below at five values of $y$ shown in colours) concentrates into steeper, larger nonlinear waves under the dynamics of the two-time CH2 equations (\ref{m-eqn}),(\ref{rho-eqn}). Figures are courtesy of L. \'O N\'araigh.  }
\label{x-profiles_figure}
\end{figure}

The first equation in this system is the spectral problem (\ref{SPCH2}) of the CH2 hierarchy. The second equation introduces the other `time' derivative, with respect to $y$.  The system (\ref{m-eqn}),(\ref{rho-eqn}) appears on setting 
\[
\left(\partial_t - \frac{1}{2\zeta} \partial_y\right)\Psi_{xx}
=
\partial_x^2 \left( \Psi_{t}-\frac{1}{2\zeta}\Psi_y\right),
\]
then using (\ref{ev-prob}),(\ref{evol-y}) to eliminate higher derivatives and assuming $\zeta_t=0=\zeta_y$.

\begin{figure}[ht]
\begin{center}
\includegraphics[width=0.32\textwidth,angle=0]{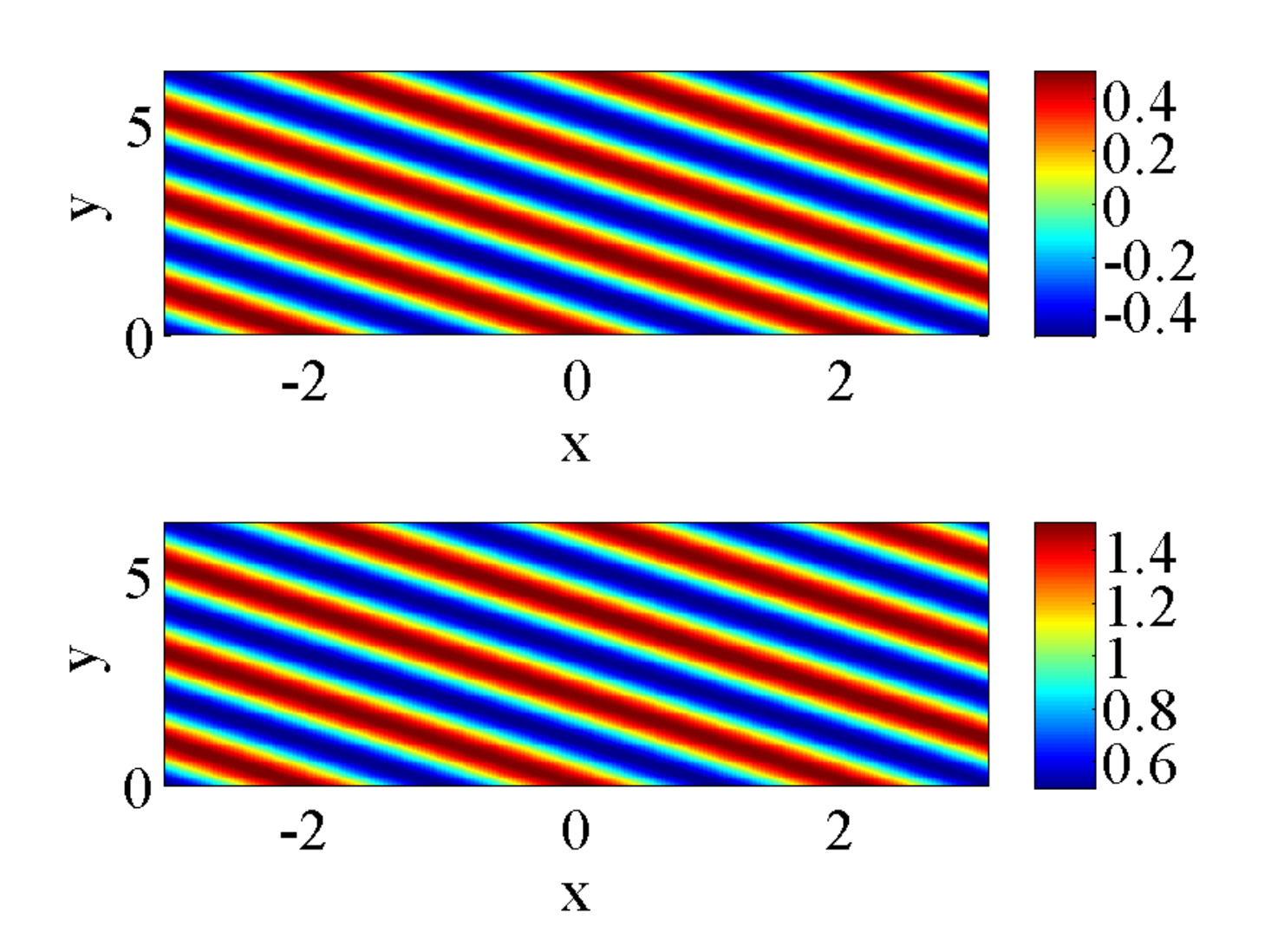}
\includegraphics[width=0.32\textwidth,angle=0]{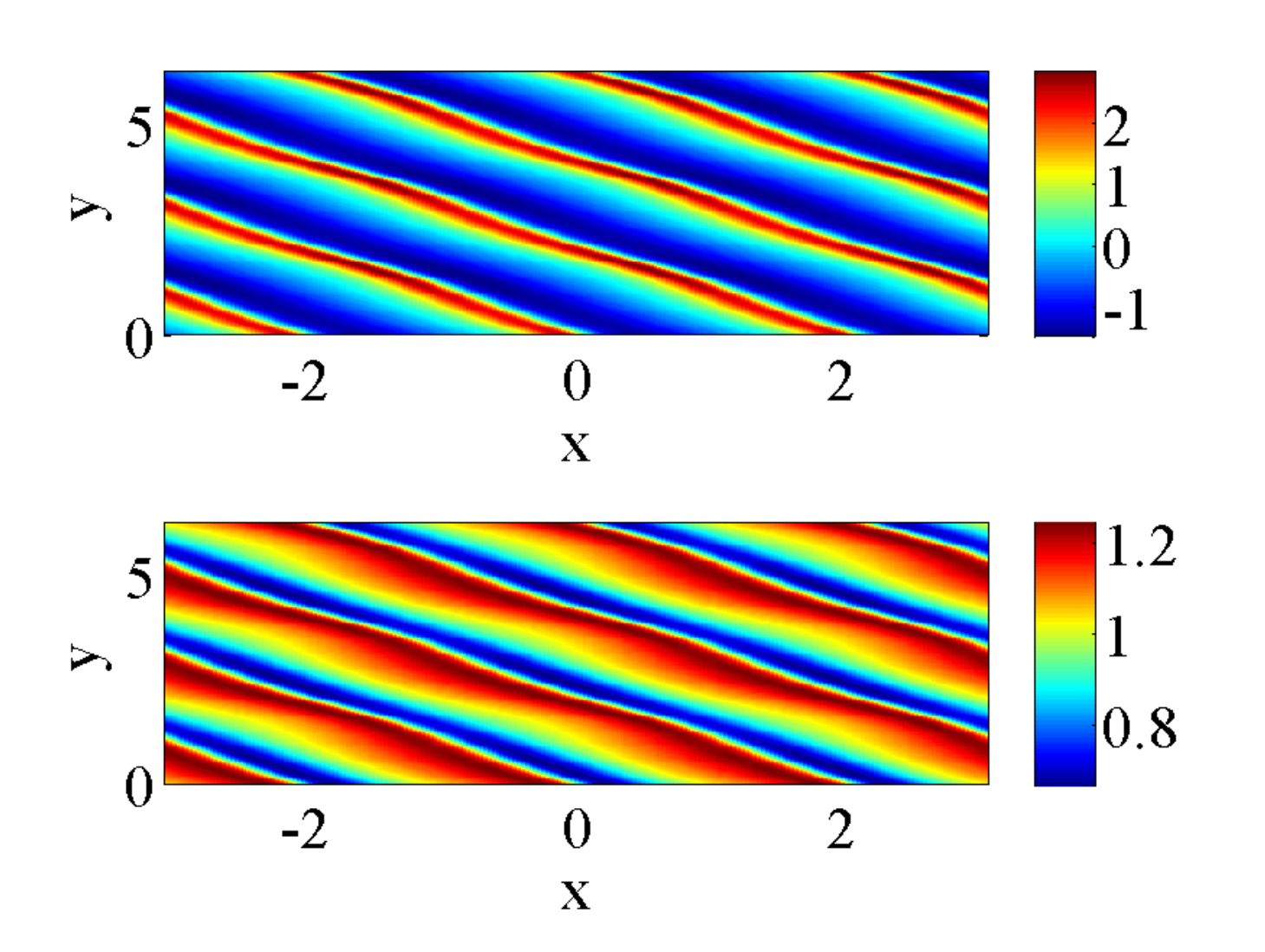}
\includegraphics[width=0.32\textwidth,angle=0]{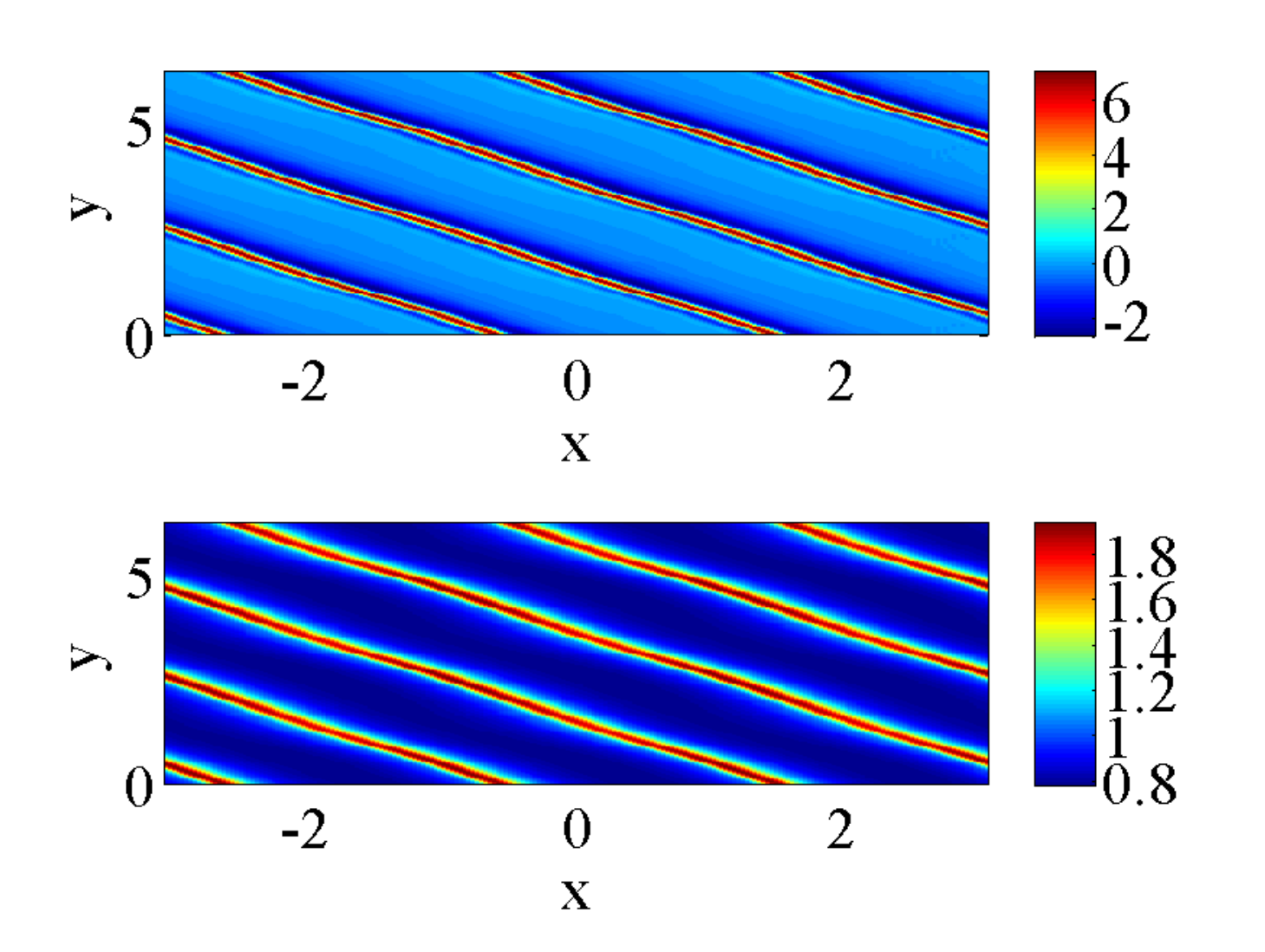}
\end{center}\vspace{-8mm}
\caption{
An initially sinusoidal wave train in the $(x,y)$ plane (whose amplitudes are shown by colour bars) with $m$ above and $\rho$ below, steepens and modulates along the crests as it grows into a sequence of nonlinear wave packets under the dynamics of the two-time CH2 equations (\ref{m-eqn}),(\ref{rho-eqn}). Figures are courtesy of L. \'O N\'araigh. 
 }
\label{xy-profiles_figure}
\end{figure}

{\bf Remarks.}
\begin{itemize}
\item
Perhaps not unexpectedly, the corresponding modification of the linear system (\ref{L1}), (\ref{L2}) yields a two-time version of the entire CH(n,k) hierarchy in (\ref{L6}).

\item
The integrable system of two-time CH2 equations (\ref{m-eqn}),(\ref{rho-eqn}) reduces to CH2 for $x=y$ and $u=U_x$. 

\item
Likewise, the special case $\gamma =0=\omega$ with initial condition $\rho=0$ admits $N$-peakon solutions, 
\begin{eqnarray} 
m(x,t,y) = \sum_{a=1}^N p_a(t,y)\,\delta(x-q_a(t,y))
\,,
\label{2Dpeakons}
\end{eqnarray} 
with two `time' variables ($t$ and $y$). 

\item
The amplitudes of $m$ and $\rho$ as functions of $x$ and $y$ in Figure \ref{xy-profiles_figure} show modulations along the crest of the two-dimensional solitons as they form under the two-time CH2 dynamics. . 
\end{itemize}

\paragraph{Graded Lie algebra structure for the CH(n,k) chain with ${\rm n}\ge3$ and ${k}>0$.}
A \emph{new feature} of the CH(n,k) chain may be recognized for ${\rm n}\ge3$ and ${k}>0$. In that case, the Hamiltonian operator reveals its character as the Lie-Poisson operator defined on the dual space of a {\bfi graded Lie algebra}. 
For example, in the case of CH(3,1) the system in the first line of (\ref{L6}) may be written in Lie-Poisson Hamiltonian form, as
\begin{equation}
\begin{bmatrix}
\partial_t q_1 \\
\partial_t q_2 \\
\partial_t q_3
\end{bmatrix}
= - 
\begin{bmatrix}
\partial_x q_1 + q_1 \partial_x & \partial_x q_2 + q_2 \partial_x & \partial_x q_3 + q_3 \partial_x \\
\partial_x q_2 + q_2 \partial_x & \partial_x q_3 + q_3 \partial_x &  0 \\
 \partial_x q_3 + q_3 \partial_x & 0 & 0
\end{bmatrix}
\begin{bmatrix}
\delta h /\delta q_1 = u_0 \\
\delta h /\delta q_2 =  - \frac12 \\
\delta h /\delta q_3 = 0
\end{bmatrix}
  .
  \label{n=3 k=1 Ham form}
\end{equation}
For $q_3=-\rho^2$ this generalizes the CH(2,1) equation to three components. In hindsight, we see that we could have written the CH(2,1) equation in the same Lie-Poisson Hamiltonian form, as 
\begin{equation}
\begin{bmatrix}
\partial_t q_1 \\
\partial_t q_2
\end{bmatrix}
= - 
\begin{bmatrix}
\partial_x q_1 + q_1 \partial_x & \partial_x q_2 + q_2 \partial_x \\
\partial_x q_2 + q_2 \partial_x &  0
\end{bmatrix}
\begin{bmatrix}
\delta h /\delta q_1 = u_0 \\
\delta h /\delta q_2 = -\frac12
\end{bmatrix}
  ,
  \label{SDP CH2 Ham form}
\end{equation}
for the Hamiltonian $h(q_1,q_2) = \frac12\int q_1(1-\partial_x^2)^{-1}q_1 - q_2\,dx$.
Likewise, in the case of CH(3,2) the system becomes, 
\begin{equation}
\begin{bmatrix}
\partial_t q_1 \\
\partial_t q_2 \\
\partial_t q_3
\end{bmatrix}
= - 
\begin{bmatrix}
\partial_x q_1 + q_1 \partial_x & \partial_x q_2 + q_2 \partial_x & \partial_x q_3 + q_3 \partial_x \\
\partial_x q_2 + q_2 \partial_x & \partial_x q_3 + q_3 \partial_x &  0 \\
 \partial_x q_3 + q_3 \partial_x & 0 & 0
\end{bmatrix}
\begin{bmatrix}
\delta h /\delta q_1 = u_0 \\
\delta h /\delta q_2 = u_1 \\
\delta h /\delta q_3 = - \frac12 
\end{bmatrix}
  .
  \label{n=3 k=2 Ham form}
\end{equation}
This grading of the Hamiltonian operator according to the weight ${\rm n}$ reveals the new feature. For CH(3,k) with ${k}>0$, the Hamiltonian operator in (\ref{n=3 k=2 Ham form}) defines a Poisson bracket on the dual of the Lie algebra of three-component vector fields $(X_1,X_2,X_3)\in \mathfrak{X}_1\times\mathfrak{X}_2\times\mathfrak{X}_3$ defined by their graded commutation relation 
\begin{equation}
[(X_1,X_2,X_3), (Y_1,Y_2,Y_3) ]
=
( [X_1,Y_1], [X_1,Y_2] + [X_2,Y_1], [X_1,Y_3] + [X_2,Y_2]  + [X_3,Y_1] ).
  \label{GradedLieAction}
\end{equation}
Dual coordinates are $q_1\in \mathfrak{X}_1^*(\mathbb{R})$, $q_2\in \mathfrak{X}_2^*(\mathbb{R})$ and  $q_3\in \mathfrak{X}_3^*(\mathbb{R})$. 
The Lie-Poisson bracket above may now be written as
\begin{eqnarray}
\Big\{
F,\,
H
\Big\}
&=&
-
\left\langle 
\Big(q_1, q_2, q_3 \Big)
\,,\,
\left[ \left(
\frac{\delta F}{\delta q_1}, 
\frac{\delta F}{\delta q_2}, 
\frac{\delta F}{\delta q_3}
\right)
\,,\,
\left(
\frac{\delta H}{\delta q_1}, 
\frac{\delta H}{\delta q_2}, 
\frac{\delta H}{\delta q_3}
\right)
\right]
\right\rangle
\,,
\end{eqnarray}
in terms of the graded commutation relation (\ref{GradedLieAction}) and the $L^2$ pairing $\langle \,\cdot\,,\,\cdot\, \rangle $ between the graded Lie algebra and its dual.

{\bf Remarks.}

$\bullet$ A Lie-Poisson bracket defined on the dual of the same graded Lie algebra also appears in plasma theory,  at third order in the Bogoliubov-Born-Green-Kirkwood-Yvon (BBGKY)  hierarchy of equations \cite{MaMoWe1984}. \smallskip 

$\bullet$ From the graded Lie-algebra action (\ref{GradedLieAction}) and the Lax pair in (\ref{L1}) - (\ref{L2}), it is clear how to extend the pattern to higher order and thus include the more deeply nested systems in the CH(n,k) chain.

$\bullet$ The semidirect product action in (\ref{sdpLieAction}) is also an instance of the graded Lie-algebra action. 

$\bullet$ This type of weighted Lie-Poisson bracket is also encountered in the  Hamiltonian formulation of dynamics in the BBGKY hierarchy for ideal plasma physics \cite{MaMoWe1984}. 

$\bullet$ A graded extension of (\ref{top-prob}) analogous to that for the Hamiltonian operator in (\ref{n=3  k=2 Ham form}) is also available for coupling with additional $so(3)^*$ angular momenta. 

$\bullet$ The properties of the full Hamiltonian structure for the CH(n,k) chain in (\ref{L6}) consisting of $n$ evolution equations with $|{k}|$ differential relations that allow ${k}<0$ will be discussed elsewhere. (For ${k}<0$, the grading runs in the `opposite direction' in a certain sense.)

\section{Conclusion} \label{conclusion-sec}

\paragraph{Main results of the paper.}
We first formulated the Lax pair consisting of the energy-dependent isospectral problem and evolution equation (\ref{L1}), (\ref{L2}), whose compatibility yields the integrable family of CH(n,k) systems (\ref{CH-nk}) with  $n$ components (momenta) and $1\le|{k}|\le n$ velocities. After looking at several examples among the CH(n,k) multi-component equations, we investigated some of the other equations of the CH2 hierarchy and found geometrical comparisons with systems of coupled spinning tops, as well as fluids because of the semidirect-product nature of their Lie-Poisson brackets. In particular, the integrable CH(2,2) equations in  (\ref{22-eq}) with Hamiltonian matrix in equation (\ref{SDP Ham form}) were seen to be analogous to the finite-dimensional equations for a spinning top in a potential force field \cite{Bog1985,{HoGM1},{HoGM2}}, whose Lie-Posson bracket is dual to the semidirect-product Lie algebra $\mathfrak{so}(3)\circledS \mathfrak{so}(3)_0$, in which the second entry is treated simply as a vector space, as discussed in \S\ref{geom(2,2)}. 

 Section  \S\ref{eqns-hierarchy-sec} provided additional examples of other integrable equations in the CH2 hierarchy, such as the CH(2,\,-1) system and the two-time version of the CH(2,1) equations. The CH(n,k) systems arising from (\ref{L1})-(\ref{L3}) with negative values of ${k}$ were found to show quite different Hamiltonian structures from their corresponding systems with positive values of ${k}$.
 
\paragraph{Properties of the CH(n,k) systems.} 
Several properties of the CH(n,k) systems were identified in the course of this work. These included the following.  
\begin{description}
\item
(i) The differential relations in the middle equation of (\ref{L6}) involved both of the compatible Poisson operators in the biHamiltonian structure (\ref{eq2}). In both CH(2,1) and CH(2,2) the differential relations defined the momenta dual to the velocity vector fields. 
\item
(ii) The Hamiltonian structure for the CH(n,k) chain with ${k}>0$ in (\ref{L6}) consisting of $n$ evolution equations and $k+1$ differential relations possesses a \emph{graded} Lie algebra structure, which became evident for ${\rm n}\ge3$. In hindsight, looking at (\ref{SDP Ham form}), the semidirect-product Lie algebra structure for $n=2$ in (\ref{SDP CH2 Ham form}) could have already been understood as being graded. For ${k}<0$ again both of the compatible Poison operators appear.
\item
(iii) The sample numerical simulations shown for the CH(2,1), CH(2,2) and CH(2,-1) equations revealed challenging properties for future investigation, such as different types of collision behavior. In the case of CH(2,1) with two times, the simulations also revealed modulation of the waves along their crests during the formation of the soliton trains. These sample numerical simulations provided insight into the fascinating pulse-like solution behavior of the CH(n,k) equations and invited further investigation.
\end{description}

\paragraph{Future challenges.} 
The CH(n,k) family of integrable partial differential equations (PDE) discussed here offer many interesting challenges for future research, particularly in determining and analyzing their solution behavior and possible physical applications. Besides CH(2,1) which may be interpreted as a shallow water system the CH(2,2) equations in \S\ref{examples CH2hierarchy}, the CH(2,-1) system in \S\ref{CH2Dym} and the two-time CH2 equations in \S\ref{2timeCHeqns} all offer new challenges for physical interpretation and mathematical analysis. For example, one may expect the continuing interest in wave-breaking analysis for CH and CH2 to extend also to the other integrable PDE in the rest of the CH(n,k) family, including, e.g., the CH(2,-2) system, which was not discussed here. The numerical simulation of these integrable PDE, and the formulation and analysis of their discrete versions can also be expected to attract attention in future endeavors. Finally, the multi-dimensional extensions and deeper geometrical aspects of these new integrable PDE also pose interesting challenges for future research. 
 
\section*{Acknowledgements}
DDH was partially supported by the Royal Society of London, Wolfson Scheme.
RII acknowledges funding from a Marie Curie Intra-European Fellowship. 
Both authors thank L. \'O N\'araigh, J. R. Percival  and V. Putkaradze for generously providing figures from their numerical solutions in ongoing investigations of the various equations in the CH(n,k) hierarchy treated here. We also thank M. Bruveris, F. Gay-Balmaz, J. Gibbons, J. E. Marsden, T. Ratiu and C. Tronci for encouraging comments and insightful remarks during the course of this work.

\end{document}